\documentclass[acmsmall]{acmart}
\AtBeginDocument{%
  }

\setcopyright{cc}
\setcctype{by}
\acmJournal{TOPS}
\acmYear{2026} \acmVolume{1} \acmNumber{1} \acmArticle{}
\acmMonth{1} \acmPrice{} \acmDOI{10.1145/3794848}


\usepackage{tikz}
\usepackage{xprintlen}
\usepackage{hyperref}
\usepackage{enumerate}

\usepackage{xcolor}
\usepackage[detect-weight=true,detect-family=true]{siunitx}
\usepackage{xspace}

\usepackage{amsmath}

\usepackage{tikz}
\usetikzlibrary{positioning}

\usepackage{multirow}

\usepackage{textgreek}

\definecolor{darkgrey}{RGB}{80,80,80}
\definecolor{lightgrey}{RGB}{170,170,170}

\definecolor{brown}{HTML}{a52a2a}
\definecolor{darkcyan}{HTML}{0a888a}

\usepackage{adjustbox}
\usepackage{booktabs}
\usepackage{wasysym}
\usepackage{tabularx} 
\usepackage{multirow}
\usepackage{arydshln} 
\usepackage{siunitx}
\usepackage{subcaption}
\usepackage{amsmath}

\newcommand{\etal}{\textit{et~al.}\xspace}
\newcommand{\ie}{\textit{i.e.},\xspace}
\newcommand{\eg}{\textit{e.g.},\xspace}
\newcommand{\cf}{\textit{cf.}\xspace}

\newcommand{\code}[1]{\texttt{#1}}

\usepackage{relsize}
\newcommand{\shortunderscore}{\textscale{.6}{\textunderscore}}

\usepackage{paralist}

\usepackage{tabularray}
\usepackage{wasysym}

\newcommand{\name}{\textsc{\mbox{Seldom}}\xspace}
\newcommand{\backref}{\textsc{\mbox{BackRef}}\xspace}

\usepackage[most]{tcolorbox}
\newtcolorbox{challengebox}[1][]{
  enhanced,
  breakable,
  arc=0pt,
  outer arc = 0pt,
  sharp corners,
  boxrule=2pt,boxsep=0pt,top=2pt,left=2pt,right=2pt,bottom=2pt,middle=-5pt,
  colback=gray!10, 
  colframe=gray!10, 
  pad at break=0pt,bottomrule at break=0pt,toprule at break=0pt,
  borderline east={0.8pt}{-0.25pt}{black,dotted},
  borderline west={0.8pt}{-0.25pt}{black,dotted},
  borderline south={0.8pt}{-0.25pt}{black,dotted},
  borderline north={0.8pt}{-0.25pt}{black,dotted},
  title={\textcolor{black}{\textbf{#1}}},
}

\usepackage[nolist]{acronym}
\begin{acronym}[SIGMAR]
        \acro{CPS}{cyber-physical system}
        \acro{DoS}{Denial of Service}
        \acro{ICS}{Industrial Control System}
        \acro{OT}{Operational Technology}
        \acro{IDS}{intrusion detection system}
        \acro{IPS}{intrusion prevention system}
        \acro{MAC}{Message Authentication Code}
        \acro{IIoT}{Industrial Internet of Things}
        \acro{TEE}{Trusted Execution Environment}
        \acro{LEA}{Law Enforcement Agency}
        \acro{CSAM}{Child Sexual Abuse Material}
        \acro{PoA}{Proof of Authority}
        \acro{SMC}{Secure Multiparty Computation}
        \acro{TPKE}{Threshold Public Key Encryption}
		\acro{ISP}{Internet Service Provider}
	\acrodefplural{LEA}{Law Enforcement Agencies}
	\acro{NFOT}{Non-Interactive Fractional Oblivious Transfer}
\end{acronym}

\newcommand{\stepnumber}[2][]{\raisebox{.5pt}{\textcircled{\raisebox{-.5pt}{\footnotesize#2\raisebox{1.5pt}{\tiny #1}}}}}

\newcommand{\stepnumberspace}[3][]{\stepnumber[#1]{#2}{\kern 0.25em}#3}

\usepackage{enumitem}

\usepackage{listings}
\definecolor{remix-blue}{HTML}{1A1AA6}
\definecolor{remix-lightblue}{HTML}{585CF6}
\definecolor{remix-magenta}{HTML}{930F80}
\definecolor{remix-green}{HTML}{236E24}
\definecolor{remix-turquoise}{HTML}{318495}

\definecolor{highlight-public}{HTML}{9dcc9b}
\definecolor{highlight-exit}{HTML}{9f81b8}
\definecolor{highlight-stats}{HTML}{e8a579}

\newcommand{\lcolorbox}[2]{%
  \hspace*{-\fboxsep}\colorbox{#1}{#2}%
}

\lstdefinelanguage{Solidity}{
	keywords=[1]{anonymous, assembly, assert, balance, break, call, callcode, case, catch, class, constant, continue, constructor, contract, debugger, default, delegatecall, delete, do, else, emit, event, experimental, export, external, false, finally, for, function, gas, if, implements, import, in, indexed, instanceof, interface, internal, is, length, library, log0, log1, log2, log3, log4, memory, modifier, new, payable, pragma, private, protected, public, pure, push, require, return, returns, revert, selfdestruct, send, solidity, storage, struct, suicide, super, switch, then, this, throw, transfer, true, try, typeof, using, value, view, while, with, addmod, ecrecover, keccak256, mulmod, ripemd160, sha256, sha3}, 
	keywordstyle=[1]\color{remix-magenta}\bfseries,
	keywords=[2]{address, bool, byte, bytes, bytes1, bytes2, bytes3, bytes4, bytes5, bytes6, bytes7, bytes8, bytes9, bytes10, bytes11, bytes12, bytes13, bytes14, bytes15, bytes16, bytes17, bytes18, bytes19, bytes20, bytes21, bytes22, bytes23, bytes24, bytes25, bytes26, bytes27, bytes28, bytes29, bytes30, bytes31, bytes32, enum, int, int8, int16, int24, int32, int40, int48, int56, int64, int72, int80, int88, int96, int104, int112, int120, int128, int136, int144, int152, int160, int168, int176, int184, int192, int200, int208, int216, int224, int232, int240, int248, int256, mapping, string, uint, uint8, uint16, uint24, uint32, uint40, uint48, uint56, uint64, uint72, uint80, uint88, uint96, uint104, uint112, uint120, uint128, uint136, uint144, uint152, uint160, uint168, uint176, uint184, uint192, uint200, uint208, uint216, uint224, uint232, uint240, uint248, uint256, var, void, ether, finney, szabo, wei, days, hours, minutes, seconds, weeks, years},	
	keywordstyle=[2]\color{remix-magenta}\bfseries,
	keywords=[3]{block, blockhash, coinbase, difficulty, gaslimit, number, timestamp, msg, data, gas, sender, sig, value, now, tx, gasprice, origi},	
	keywordstyle=[3]\color{remix-blue}\bfseries,
	comment=[l]{//},
	morecomment=[s]{/*}{*/},
	commentstyle=\color{remix-green}\ttfamily,
	stringstyle=\color{remix-blue}\ttfamily,
	morestring=[b]',
	morestring=[b]"
}

\begin{document}

\title{\name: An Anonymity Network with Selective Deanonymization}

\author{Eric Wagner}
\orcid{0000-0003-3211-1015}
\affiliation{
  \institution{Fraunhofer FKIE}
  \country{Germany}
}
\affiliation{
  \institution{University of Luxembourg}
  \country{Luxembourg}
}
\email{eric.wagner@uni.lu}

\author{Roman Matzutt}
\orcid{0000-0002-4263-5317}
\affiliation{
  \institution{Fraunhofer FIT}
  \country{Germany}
}
\email{roman.matzutt@fit.fraunhofer.de}

\author{Martin Henze}
\orcid{0000-0001-8717-2523}
\affiliation{
  \institution{RWTH Aachen University}
  \country{Germany}
}
\affiliation{
  \institution{Fraunhofer FKIE}
  \country{Germany}
}
\email{henze@spice.rwth-aachen.de}

\begin{abstract}

While anonymity networks such as Tor provide invaluable privacy guarantees to society, they also enable all kinds of criminal activities.
Consequently, many blameless citizens shy away from protecting their privacy using such technology for fear of being associated with criminals.
To grasp the potential for alternative privacy protection for those users, we design \name{}, an anonymity network with integrated selective deanonymization that disincentivizes criminal activity.
\name enables law enforcement agencies to \emph{selectively} access otherwise anonymized identities of misbehaving users while providing technical guarantees preventing these access rights from being misused.
\name further ensures \emph{translucency}, as each access request is approved by a trustworthy consortium of impartial entities and eventually disclosed to the public (without interfering with ongoing investigations). 
To demonstrate \name{}'s feasibility and applicability, we base our implementation on Tor, the most widely used anonymity network.
Our evaluation indicates minimal latency, processing, and bandwidth overheads compared to Tor; \name{}'s main costs stem from storing flow records and encrypted identities.
With at most \SI{636}{TB} of storage required in total to retain the encrypted identifiers of a Tor-sized network for two years,
\name{} provides a practical and deployable technical solution to the inherent problem of criminal activities in anonymity networks.
As such, \name{} sheds new light on the potentials and limitations when integrating selective deanonymization into anonymity networks.

\end{abstract}


\maketitle

\section{Introduction}

Tor \cite{2004_dingledine_tor} and other anonymity networks provide invaluable services to, \eg journalists, whistleblowers, and militaries, who are among its millions of daily users~\cite{tor-users}.
One may argue that Tor anonymizes too well, as criminals misuse Tor to establish a hotspot of highly illegal activities, such as sharing \ac{CSAM}~\cite{2024_nurmi_investigating}.
Reports showing that over \SI{10}{\percent} of Ahmia.fi search sessions seek \ac{CSAM}~\cite{2024_nurmi_investigating} and over \SI{50}{\percent} of marketplace listings are related to illegal drugs~\cite{2023_hiramoto_illicit} underpin the need for reevaluating the societal downsides of \emph{unconditionally} anonymous Internet use.

Collectively, privacy advocates, such as researchers and some governmental agencies, may accept this fact as a necessary evil for the irreplaceable service Tor provides to society.
On an individual level, however, people do have moral concerns regarding such misuse of Tor.
Studies highlight that Tor users are concerned about illegal activities within the network~\cite{2018_winter_tor,2017_gallagher_new}.
Moreover, a study suggests that \SI{61}{\percent} of US citizens ``would like to do more'' to improve their privacy, while only \SI{2}{\percent} of participants have used anonymity networks such as Tor in the past~\cite{2015_rainie_americans}.
Consequently, there exists a potentially large untapped user base for anonymity services.
We argue that some of these users, potentially with only a passing understanding of Tor, shy away from it due to its close association with illegal activities they intend to stay away from. 
After all, each user indirectly supports all Tor activities---by increasing everyone's anonymity~\cite{2006_dingledine_anonymity-loves-company}---and (exit) relay operators become active targets of criminal investigations.

With this paper, we strive to investigate the potential advantages and limitations of an alternative anonymity network that offers integrated \textbf{sel}ective \textbf{d}ean\textbf{o}ny\textbf{m}ization of suspicious users (\name).
\name empowers \acp{LEA} to effectively prosecute crime by investigating suspicious activities within the anonymity network only with acknowledged due cause and under public oversight, while technically limiting these capabilities to prevent abuse of power at the same time.
\name thus provides those who do not want to be associated with Tor an alternative to adequately protect their anonymity on the Internet.
As the provided anonymity grows with the number of users~\cite{2006_dingledine_anonymity-loves-company}, wide acceptance of \name can offer strong anonymity even when \acp{LEA} can deanonymize selected suspicious identities based on publicly agreed-upon rules.

\name realizes its functionality to deanonymize a fraction of communication flows based on a novel \emph{oblivious authentication protocol} that links all outbound traffic to threshold-encrypted user identities.
A trustworthy consortium of impartial parties then decides about \acp{LEA}' deanonymization requests for individual communication flows.
To provide public transparency without impeding ongoing investigations, we propose an immutable but delayed public release of all activities via a \emph{translucent ledger}.
Comparing the performance of \name to Tor, we measure only an imperceptible latency and processing overhead.
\name would thus offer a Tor-like experience without the downside of supporting crime when merely using the network.
If significantly outgrowing Tor, \name could even reduce network latency and thus address one of the major drawbacks of onion routing.

\textbf{Contributions.} To provide a thorough understanding of the potential \emph{and} implications of selective deanonymization in anonymity networks, we make the following contributions:
\begin{itemize}[topsep=5pt,leftmargin=15pt]
        \item We devise an oblivious authentication protocol that links each outgoing communication flow to a unique threshold-encrypted identity of the anonymized client, and we integrate this protocol into our anonymity network \name.
        \item We design a translucent deanonymization process that only reveals the client identity to the requesting \ac{LEA}.
        \item Our evaluation shows that \name has an imperceptible performance impact on the users' experience and imposes only well-manageable data storage requirements on centralized databases.
        \item Finally, we discuss the challenges, limitations, and implications of deploying an anonymity network with exceptional deanonymization capabilities.
\end{itemize}

\section{Acceptable Exceptional Access}
\label{sec:scenario:properties}

Providing data access to an \ac*{LEA} in legitimate circumstances is often referred to as \emph{exceptional access}.
Any system providing exceptional access naturally faces public scrutiny due to its potential for abuse.
It is thus vital to technologically protect individual privacy against \acp{LEA} by enforcing regulation, transparency, and accountability.
However, \acp{LEA} need to be able to identify unfolding crimes and act accordingly.  
To combine these conflicting requirements, we lay out the prerequisites to successfully strike a balance in the safety-privacy dilemma.
Overall, we identify the following aspects to be essential for a widely acceptable form of exceptional access:


\textbf{Nobody But Us.}
%
Exceptional access must be restricted to authorized entities and must especially not be accessible to criminals or privacy-invasive third parties.



\textbf{Translucency.}
\acp{LEA} should operate transparently and inform the public of targeted surveillance activities.
However, timely access to such data also helps criminals evade surveillance. 
Hence, \emph{translucency}, which allows for critical information to remain \emph{temporarily} classified, is needed.


\textbf{Accountability.}
If an \ac{LEA} abuses its given privileges, its exceptional access privileges must be revocable in a way that is transparent to the system's users.

\textbf{Ethical Decision-Making.}
Individual \acp{LEA} might target different persons of interest.
For instance, government-critical journalists or whistleblowers might be subjected to surveillance without posing a harm for public safety~\cite{trump-whistleblowing}.
Any decision-making process on how to grant an \ac{LEA} exceptional access must thus reflect this ethical issue and allow for denying access on these grounds.


\textbf{Timely and Flexible Access.}
\acp{LEA} must be able to swiftly access data in case of imminent danger.

\textbf{Limited User Overhead.}
When using a system offering exceptional access, citizens should not be unnecessarily affected by overhead stemming from enabling this access.

Overall, these essential prerequisites ought to ensure that surveillance of digital communication
\begin{inparaenum}[(i)]
    \item is limited to exceptional cases,
    \item is effective in prosecuting crime on the network and thereby disincentivizing it, and
    \item generates wide societal acceptance by strengthening public safety.
\end{inparaenum}

\section{Related Work}%
\label{sec:scenario:related-work}

In the following, we give an overview of previous approaches to enabling exceptional access.
We notice that related work has evolved and considerably broadened its scope since the initial proposals during the 1990s crypto wars.

\textbf{Partial Key Escrow.}
Early works focused on \emph{(partial) key escrow}, which involves the government maintaining a global store of (parts of) all used encryption keys~\cite{1996_denning_taxonomy_pke,1997_bellare_vpke}.
Hence, the government would be able to retrieve keys and brute-force partial keys if needed.
However, such systems require strong trust in key holders; even the protection of partial keys decays over time as computation power rises~\cite{2015_abelson_keys_under_doormats2}.
Similarly, the idea of obliviously escrowing encryption keys at random peers~\cite{1996_blaze_oblivious_key_escrow} was quickly abandoned due to overhead and trust concerns.
Recent work looks at the use of distributed ledgers as trusted parties for key escrow~\cite{2021_green_abuse,2023_fetzer_universally}.
These systems, however, require active adoption from users, while it remains unclear how users with criminal intent would be motivated to use such key escrow systems.

\textbf{Translucent Cryptography.}
Orthogonally to key escrow schemes, \emph{translucent cryptography}~\cite{1999_bellare_translucent_crypto} is based on a variant of oblivious transfers~\cite{1981_rabin_oblivious-transfer} called \acfp{NFOT}.
In this scheme, any message also holds an \ac{NFOT}, which allows an \ac{LEA} to decrypt wiretapped messages with a predefined probability.
However, at least one communication partner must remain honest in this scheme.
Further, evaluating a wiretapped \ac{NFOT} leaves no evidence, which prevents transparency or accountability.
Finally, a purely probabilistic per-message scheme does not hold up with today's communication patterns and surveillance capabilities.
Namely, being able to decrypt only a tiny fraction of messages still allows one to peek into a large fraction of conversations but prohibits gaining all relevant information from any rightfully investigated individual.

\textbf{Crypto Crumple Zones.} 
As a continuation of the idea of partial key escrow, \emph{crypto crumple zones}~\cite{2018_wright_crumplezone} are designed 
to augment the generation of ephemeral keys with exceptional access in mind.
At their core, they enable powerful entities to invest computational resources to retrieve full keys akin to Bitcoin's proof of work.
Overall, this approach has improved rate-limiting guarantees compared to partial key escrow, but it neither restricts access to \acp{LEA} nor provides transparency.

\textbf{Lawful Device Access.}
Another proposal considers the design of special storage hardware that can be decrypted based on prolonged physical access, \eg hours to days~\cite{2018_savage_selfescrow}.
\acp{LEA} would then have to prove continual physical access before being able to read the storage.
This way, transparency, accountability, and rate-limited access are established by requiring traditional seizures of hardware.
Namely, this approach could have provided a legitimate framework for accessing the device at the center of the controversy around the San~Bernardino shooting~\cite{apple_vs_fbi}.
However, the roll-out of such technology would be slow due to its reliance on the wide adoption of specialized storage hardware, and it does not consider encrypted containers stored on such hardware.

\textbf{Privacy-preserving Content Scanning.}
Orthogonally, client-side scanning wants to identify illegal material such as \ac{CSAM} on clients' devices.
In this vein, \emph{privacy-preserving content scanning} attempts to realize client-side scanning while respecting the user's privacy~\cite{2021_kulshrestha_identifying, 2023_bartusek_end}.
However, current approaches suffer from limitations, \eg the uncertainty of whether detection remains restricted to harmful material and will not be abused for censorship or surveillance of political opponents~\cite{2024_abelson_bugs}.
Indeed, recent analysis shows how the underlying perceptual hashing algorithms to compare images are susceptible to adversarial images, \ie images in the database that look like \ac{CSAM} to humans but have hashes that collide with other secretly tracked content~\cite{2022_jain_adversarial, 2023_jain_deep, 2024_hooda_experimental}.

\textbf{Deanonymization in Onion Routing.}
A wide range of approaches strives to enable deanonymizations of activities in anonymity networks through traffic analysis, such as website fingerprinting on the connection between client and entry relay, \eg to derive which website a client likely visited.
To this end, different approaches claim to identify visited websites based on traditional machine learning \cite{panchenko2011website,wang2014effective,hayes2016kfingerprinting,dyer2012peek,cai2012touching,panchenko2016website}, deep learning \cite{rimmer2018automated,sirinam2019triplet,shen2023subverting,rahman2020tik,bhat2019var}, or generative artificial intelligence \cite{oh2021gandalf}.
However, besides various issues with website fingerprinting itself, such as unrealistic or oversimplified assumptions~\cite{juarez2014critical} or limited scalability~\cite{panchenko2016website,cherubin2022online}, these approaches neither provide correctness guarantees nor allow deanonymizing clients when observing activity related to the server side, \eg who accessed a certain website.
To the best of our knowledge, \backref~\cite{2014_backes_backref} is the only proposal to provide such functionality by enabling relays to iteratively prove that they are not the origin of suspicious traffic.
\backref~\cite{2014_backes_backref}, however, creates significant storage overhead for all relays and requires them to prove their innocence, which becomes problematic if relay operators cannot provide evidence, \eg due to data loss.
Finally, \backref~\cite{2014_backes_backref} is neither transparent in terms of scale nor purpose of requests. 

Past work on exceptional access thus aims to provide \acp{LEA} with tools to intrude on an individual's privacy to ultimately protect public safety.
In contrast, we are the first to propose using exceptional access capabilities to increase privacy by providing a new  service to an untapped user base currently browsing the Internet without anonymity.

\section{High-Level Overview of \name}%
\label{sec:design}

In this section, we give a high-level overview of \name's design and the assumed threat model before we describe \name's technical details in greater depth in the following Sections~\ref{sec:authentication}~and~\ref{sec:deanonymization}.

\subsection{Design Overview}

\begin{figure}

  \centering
  \begin{subfigure}[b]{\textwidth}
      \centering
      \includegraphics[width=.95\columnwidth,trim={0 3.0cm 0 0},clip,page=2]{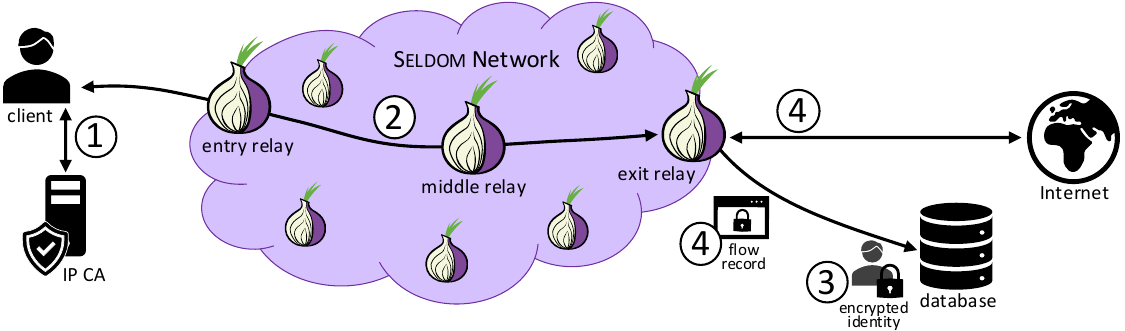}
      \caption{
        To anonymize their identity in \name, a client first \stepnumberspace{1}{requests} an IP~certificate before \stepnumberspace{2}{establishing} a new circuit.
        The exit relay \stepnumberspace{3}{outsources} encrypted identities before \stepnumberspace{4}{connecting} the client to the Internet and storing flow records in the database.
      }
      \label{fig:overview:anonymization}
  \end{subfigure}

  \begin{subfigure}[b]{\textwidth}
      \centering
      \includegraphics[width=.95\columnwidth,trim={0 2.8cm 0 0},clip,page=3]{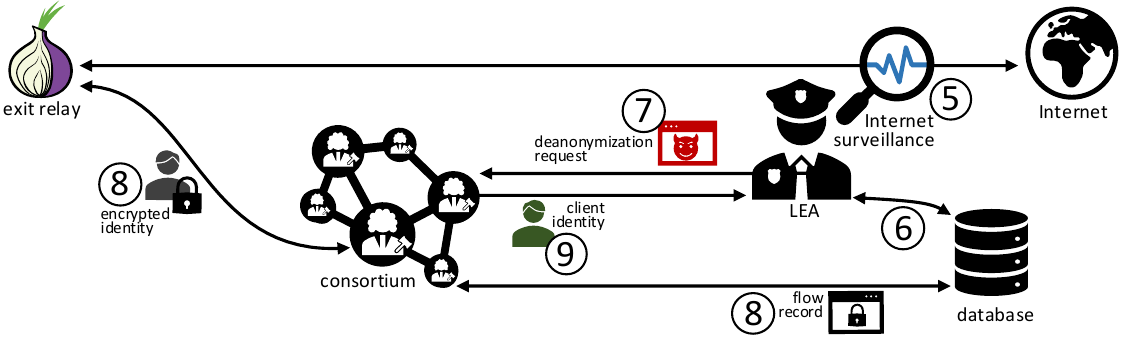}
      \caption{
        \acp{LEA} monitor the Internet and, upon \stepnumberspace{5}{identifying} suspicious traffic, an \ac{LEA} can \stepnumberspace{6}{look up} the corresponding flow record and \stepnumberspace{7}{request} a deanonymization.
        Then, the consortium \stepnumberspace{8}{obtains} the required information and votes to \stepnumberspace{9}{reveal} the client's identity to the \ac{LEA}.
      }
      \label{fig:overview:deanonymization}
  \end{subfigure}

    \caption{
        Overview of \name's anonymization (Figure~\ref{fig:overview:anonymization}) and deanonymization (Figure~\ref{fig:overview:deanonymization}) procedures.
    }
	\label{fig:overview}
\end{figure}

\name is an anonymity network based on onion routing that is extended with exceptional access capabilities according to our prerequisites formulated in Section~\ref{sec:scenario:properties}.
We design \name by extending Tor's design.
Figure~\ref{fig:overview} provides a high-level overview of \name's architecture.
Like Tor, the \name network uses an ensemble of directory authorities to manage the relays in the network and also supports bridge relays that do not need to be listed publicly. 

Clients use \name just like Tor, but \name securely stores data related to the clients' identities in a database under shared control of a \emph{consortium} of impartial parties. 
This process allows an \ac{LEA} observing suspicious Internet traffic leaving the \name network to file a substantiated \emph{deanonymization request} with the consortium.
It is the consortium that then \emph{jointly} decides whether it should reveal the client's identity to the requesting \ac{LEA}, with the knowledge that this decision and the attached justification will be released to the public in the future.

The anonymization process of \name is illustrated in Figure~\ref{fig:overview:anonymization}.
First, the client must \stepnumberspace{1}{obtain} a certificate based on their current IP~address as a means for potential identification.
To then \stepnumberspace{2}{establish} a circuit, the client propagates the IP~certificate to the exit relay such that all relays can \emph{obliviously validate} its authenticity.
At the end of the circuit establishment, the exit relay \stepnumberspace{3}{stores} the encrypted client identity in a, for now, government-sponsored central database. 
These identities are only threshold-decryptable by the consortium.
Afterward, the circuit is established successfully, and the client can \stepnumberspace{4}{access} the Internet.
Simultaneously, the exit relay \stepnumberspace{4}{stores} symmetrically encrypted flow records with searchable metadata for each connection in the database. 

We assume that \acp{LEA} are monitoring the Internet.
If they \stepnumberspace{5}{identify} suspicious activity coming from the \name network, they proceed as shown in Figure~\ref{fig:overview:deanonymization}.
They first \stepnumberspace{6}{query} the database for the corresponding flow record and \stepnumberspace{7}{file} a duly substantiated deanonymization request with the consortium.
The consortium logs the request and potential authorization through a smart contract on a translucent ledger.
After receiving proof of a logged and approved deanonymization request, the exit relay \stepnumberspace{8}{shares} the corresponding circuit's \emph{encrypted} identity with the consortium.
Then, the consortium \stepnumberspace{9}{decrypts} the threshold-encrypted identity and reveals the identity only to the requesting \ac{LEA}.

Immutably logging \emph{all} requests and decisions is crucial for maintaining transparency toward the users.
However, \name has to ensure that criminals under investigation do not gain an advantage via this channel.
To address this issue, we introduce a \emph{translucent ledger} (\cf~Section~\ref{sec:recovery:ledger}) that is maintained by the consortium and allows for a \emph{delayed} full disclosure while providing a general overview of ongoing activities in real time.
With this form of public oversight, citizens can notice \emph{patterns} of potential power abuses 
within \name and abandon the network in cases of government overreach.
In case of a successful deanonymization, users also only lose exactly the anonymity gained through the use of the network, \ie their browsing is as safe or unsafe as if no anonymity network was ever used.

\subsection{Threat Model and Trust Assumptions}
\label{sec:threat_model_and_assumptions}

The threat model for \name and the trust assumptions toward the different entities involved are guided by the core goal of \name:

\begin{challengebox}[Mission Statement of \name]
  \name's purpose is to provide a crime-free anonymization network to encourage more users to use such networks.
  To deter malicious activities, the anonymity provided by \name is revocable in justified cases, \eg if public safety is at risk.
  This revocation power is protected against abuse from any involved entity by relying on strong cryptographic primitives that only enable policy-based and translucent deanonymization.
  Further, deanonymization is non-invasive in that it is strictly limited to the service provided by \name.
  It is an explicit goal that a client establishing a TLS connection through \name can be deanonymized while the transmitted traffic remains encrypted.
  If backdoored encryption should ever be desired, it falls outside of \name's scope. 
\end{challengebox}

In the following, we first introduce our threat model before discussing the trust assumptions toward the different entities in the network.

\subsubsection{Threat Model}
\label{sec:threat_model}


We build on Tor's threat model~\cite{2004_dingledine_tor} and extend it to capture the deanonymization capabilities introduced by \name.
Hence, we assume that there exists no global passive adversary who is capable of matching all incoming and outgoing traffic at every relay.
Instead, we consider adversaries that can observe a fraction of all Internet traffic and possibly operate a portion of relays.
This direct passive observation is one of the ways for \acp{LEA} to identify critical traffic on the Internet that may require deanonymization.
Alternatively, \acp{LEA} can request (usually through a warrant) traffic metadata, \eg IP addresses and timestamps, from \acp{ISP} or website operators.
While a single person or entity can take on multiple roles in the network, it cannot take over a significant share of \eg relay operators.
An \ac{LEA} can, for example, take on the role as client and relay operator in the network at the same time.
Further, a government, as one entity, could even act as an \ac{LEA} and a consortium member.
With these capabilities, the goals of an adversary can be twofold.
They can either evade or prevent deanonymizations to hide suspicious activities, frame an innocent client, or simply disrupt the network to hinder adoption.
On the other hand, an adversary can also attack \name to conduct unauthorized deanonymizations.

\subsubsection{Trust Assumptions about the Different Roles}
\label{sec:trust_assumptions}

In the following, we list the entities involved in the operation of \name and which trust assumptions are made about them.  

\vspace*{2mm}
\noindent\textbf{Client.}
Anyone can operate one or multiple \name clients, but a client only has access to information about itself. 

\vspace*{2mm}
\noindent\textbf{IP Certificate Authority.}
IP CAs are independent entities that have to be designated when \name is set up. 
Deanonymization results of \name can only be trusted by the requesting \ac{LEA} to the degree they trust the CA.

\vspace*{2mm}
\noindent\textbf{\name Relay.}
Any entity can operate one or multiple relays.
As for Tor, clients need to trust that no single entity controls a large fraction of all relays, as proper anonymization is not guaranteed otherwise.
Relays are expected to follow the \name protocol and will be excluded from the network if misbehavior is detected. 

\vspace*{2mm}
\noindent\textbf{\name Exit Relay.}
Exit relays are relays that are exposed to the Internet and thus the initial suspicion points for \acp{LEA}.
In Tor, this risk deters many relay operators. On the other hand, \name, with its automated deanonymization process, can better protect exit relays.
In turn, exit relays are expected to be available for deanonymizations. 

\vspace*{2mm}
\noindent\textbf{Directory Authority.}
\name, like Tor, has a set of highly trusted directory authorities that manage the list of relays.

\vspace*{2mm}
\noindent\textbf{Consortium.}
The members of the consortium must be carefully chosen, and only a small number of changes over time are expected.
It must collectively be trusted to act honestly.
For \name to function properly, the consortium thus has to be trusted by clients, relays, and \acp{LEA} alike, while each individual may distrust different members within the consortium.

\vspace*{2mm}
\noindent\textbf{Database Operator.}
We assume that all entities trust the database operator(s). 
The involved risk is that they can see the metadata of all outgoing connections through flow records.
As detailed later in Section~\ref{sec:handshake:traffic} (Figure~\ref{fig:connection-info}), database operators do not see which flows are associated with the same circuit.
In practice, the database could be operated by the directory authorities, by governments to redistribute costs if the privacy implications are accepted, or distributed among multiple entities. 

\section{Oblivious Authentication}
\label{sec:authentication}

\name is built upon an \emph{oblivious authentication protocol} integrated into onion routing.
This protocol authenticates a client to an exit relay without revealing its identity.
Clients have to obtain \emph{temporal IP~certificates} (Section~\ref{sec:client-cert}) before being able to establish a circuit (Section~\ref{sec:handshake}) so that exit relays are convinced that any outgoing traffic can be irrefutably linked to the client (Section~\ref{sec:handshake:traffic}).
In Appendix~\ref{app:extension}, we detail how we integrate this protocol with the existing Tor handshake. 

\subsection{Temporal IP Certificates}
\label{sec:client-cert}

\name needs a way to reliably represent an individual's identity if deanonymization is required later on. 
Regular, CA-issued certificates are not applicable to \name, as
\begin{inparaenum}[(i)]
    \item obtaining them is costly and not always possible~\cite{2014_backes_backref} and
    \item they would share excessive personal information with entry relays.
\end{inparaenum}
Instead, we propose to identify users via \emph{temporal IP~certificates}, which link an ephemeral public key\footnote{\name uses the ed25519 signature scheme for its speed and small signatures.} of a client to a timestamped IP~address.\footnote{Since designing and implementing \name, Let's Encrypt announced the rollout of IP Address Certificate~\cite{ip_cert}, which would satisfy \name's requirements.}
While not necessarily allowing recovery of a criminal's true identity directly, deanonymizable temporal IP~certificates effectively cancel out the added privacy provided by the anonymity network only for \acp{LEA} bringing forward good reasons for further investigation. 
Only the entry relay obtains the temporal IP~certificate in the clear for validation purposes.
If the deanonymization reveals an invalid certificate, this serves as an indication for \acp{LEA} that the client and entry relay colluded, pointing to the entry relay for further investigation.

Temporal IP~certificates are issued in an automated manner by one or more CAs that are trusted by \acp{LEA}.
The client generates a public-private key pair for any new Internet connection (\eg ISP-triggered reassignment of IPv4~addresses) and sends the public key to a CA for certification.
In the simplest case, the CA extracts the client's IP~address from the request and creates the temporal IP~certificate from that IP~address, the public key, and the current timestamp.
We further discuss adaptations to this process for users relying on bridge relays, \eg for censorship evasion, in Appendix~\ref{app:ipcert}.

\subsection{Oblivious Authentication Protocol}
\label{sec:handshake}

\begin{figure}
	\centering
	\includegraphics[angle=0,width=\columnwidth]{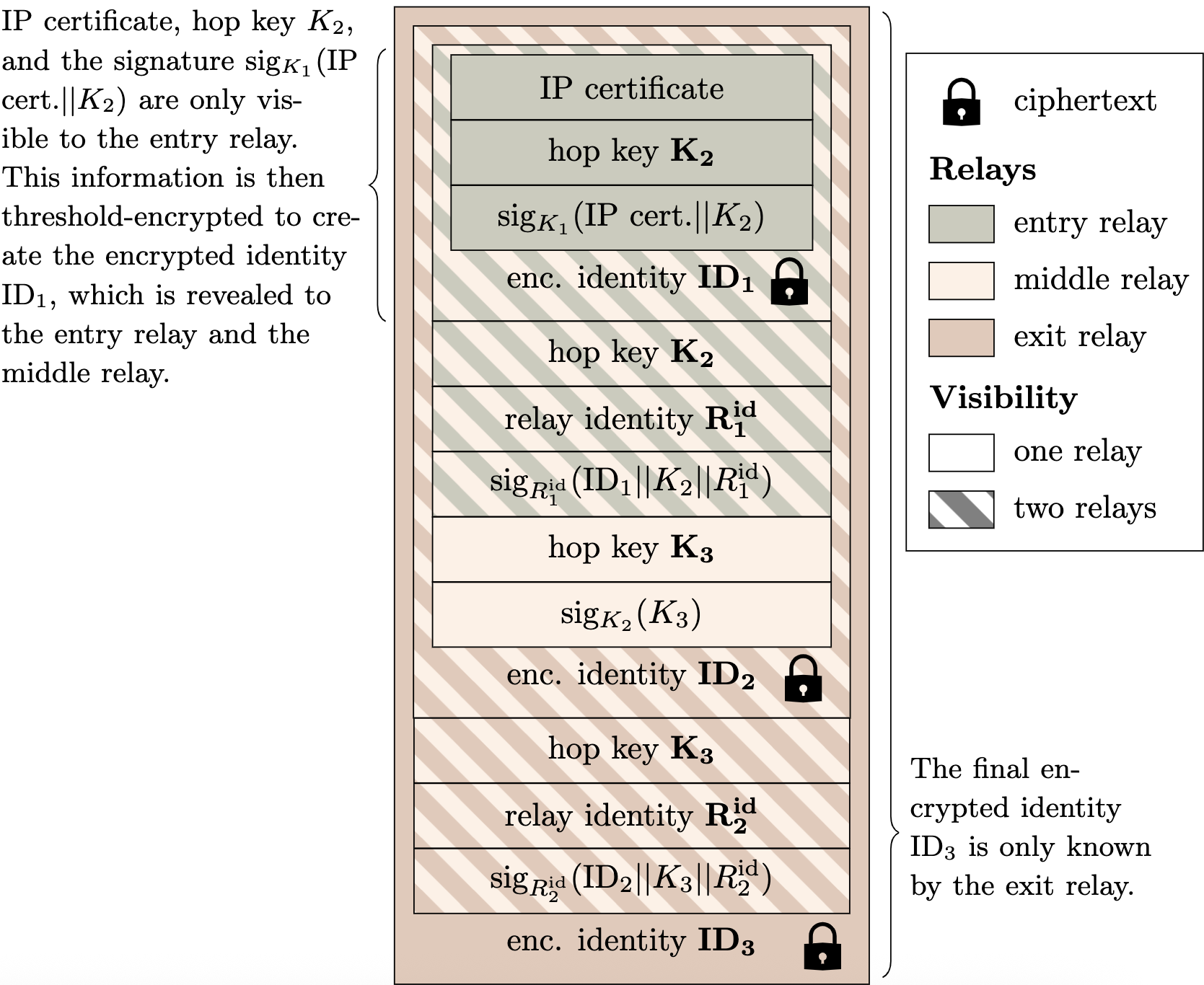}
	\caption{
		\name's oblivious authentication protocol hides the client's identity behind multiple layers of threshold encryption (entries marked with a padlock), which can only be decrypted through collaboration of the consortium. The encrypted identity of a 3-hop circuit is approximately \SI{500}{\byte} long.}
	\label{fig:encrypted-identity}
\end{figure}

We now present our oblivious authentication protocol.
This protocol distinguishes three cases, detailed in the following: authenticating the client to their entry relay (Section~\ref{sec:handshake:client-entry}), propagating authentication via middle relays (Section~\ref{sec:handshake:middle}), and finalizing authentication at the exit relay (Section~\ref{sec:handshake:exit}).
Figure~\ref{fig:encrypted-identity} illustrates the final derived encrypted identity for a circuit consisting of three relays $R_1$, $R_2$, and $R_3$.

\subsubsection{Authentication to the Entry Relay}%
\label{sec:handshake:client-entry}

The purpose of our authentication protocol is to bind a temporal IP~certificate to the hops within a circuit in a way that relays can only obtain information related to their direct predecessor.
We hence refer to the ed25519 public keys to authenticate individual hops as \emph{hop keys} $K_n$.

After obtaining a valid temporal IP~certificate~(\cf~Section~\ref{sec:client-cert}), the client confidentially sends that certificate to the entry relay $R_1$, which extracts the client's first hop key $K_1$ from the certificate.
The entry relay validates $K_1$ by verifying the IP certificate's signature by the CA and ensuring that the client uses the certified IP~address.
Failures to validate the IP~address will surface during deanonymizations; hence, $R_1$ is incentivized to perform this check honestly.
Next, the client generates a second hop key pair (with the associated public key $K_2$) to prepare the authentication forwarding between relays $R_1$ and $R_2$.
By signing $K_2$ using the validated $K_1$, $R_1$ can in turn validate the authenticity of $K_2$.
Finally, $R_1$ threshold-encrypts the client's IP~certificate, $K_2$, and the signature under $K_1$ using a CCA-secure \ac{TPKE} scheme.
Thus, only the consortium can access the encrypted data from now on.
We refer to the resulting ciphertext as the client's \emph{encrypted identity $\text{ID}_1$}.
Our concrete implementation relies on a hybrid threshold-encryption scheme based on Shoup's threshold RSA signatures~\cite{2000_shoup_threshold-rsa} and AES in GCM mode. 
We published the scheme as open source under the name \emph{thRSAhold} and present it in more detail in Appendix~\ref{app:encryption}.


\subsubsection{Oblivious Authentication Forwarding}%
\label{sec:handshake:middle}

Assuming that an obliviously authenticated partial circuit has been established up to relay $R_n$, we now describe how to extend the authentication via another hop to $R_{n+1}$.
On a high level, $R_n$ forwards its view on the state of current authentication, $\text{ID}_n$ and $K_{n+1}$ as well as its own relay identity key $R_n^{\text{id}}$ to $R_{n+1}$.
$R_n^{\text{id}}$ is usually an ed25519 public key and published via the relay's descriptor~\cite{2004_dingledine_tor}.
Other relays, such as unlisted bridge relays, can be identified pseudonymously via additional, \name-specific certificates  retrieved from a bridge authority.
Only in cases where \acp{LEA} must identify a relay operator, bridge authorities would reveal the certificate owner.
$R_n$ further signs the transmitted information with its identity key, yielding the signature $\text{sig}_{R_n^{\text{id}}}(\text{ID}_n||K_{n+1}||R_n^{\text{id}})$.
This signature is transferred to, and verified by, $R_{n+1}$.
Verifying the identity of $R_n$ reassures $R_{n+1}$ that \acp{LEA} can discover $R_n$ if necessary, and thus any deanonymization process can be properly delegated to $R_n$.

Following this step, the client establishes a new hop key $K_{n+2}$ between itself and $R_{n+1}$ without revealing it to $R_{n}$.
To link this new hop key $K_{n+2}$ to the circuit, the client signs this new hop key with the old hop key $K_{n+1}$ and sends the new key and this signature, \ie $\text{sig}_{K_{n+1}}(K_{n+2})$, to $R_{n+1}$.
This signature is the only link between two hop keys.
To prevent collusion attacks between relays, it is, therefore, crucial that this signature is only revealed to the relay also knowing the corresponding keys, \ie $R_{n+1}$. 
After validating the signature, $R_{n+1}$ computes $\text{ID}_{n+1}$ by threshold-encrypting all now-established data.

These steps can be repeated to establish longer circuits.
Only the final extension step, to the exit relay, has to be treated differently, as we detail next.

\subsubsection{Oblivious Authentication to the Exit Relay}%
\label{sec:handshake:exit}

In the last step, the exit relay obliviously authenticates the client.
Without loss of generality, we now assume that a circuit has length three, \ie the exit relay is $R_3$.
Then, $R_3$ receives the following from $R_2$: the current encrypted identity $\text{ID}_2$, the hop key $K_3$, the middle relay's identity $R^{\text{id}}_2$, and its signature $\text{sig}_{R_2^{\text{id}}}(\text{ID}_2||K_3||R_2^{\text{id}})$. 
As $R_3$ is the last hop within the circuit, it only has to threshold-encrypt the received data one last time to yield the final encrypted identity $\text{ID}_3$.
In case of future deanonymization, a client-circuit binding is identified by the last hop key $K_3$, which we also call the \emph{client identification key}, and a digest $H(\text{ID}_3)$ of the encrypted identity (\name uses SHA-3 as its cryptographic hash algorithm $H$).

However, storing available encrypted identities and associated metadata of outgoing traffic flows quickly becomes expensive for the exit relay.
Thus, \name outsources this data to external databases.
The threshold encryption of $\text{ID}_3$ prevents unlawful access by individual parties and outsourcing this information unburdens exit relays and ensures data availability to \acp{LEA} at the same time.
Therefore, exit relays are expected to upload their data periodically in batches to mitigate traffic analysis attacks and improve bandwidth usage.
The uploaded data is retained only for a predetermined period, \eg for one year.
Afterward, the consortium and exit relays refuse to support deanonymizations.


\subsection{Linking Internet Traffic to Circuits}
\label{sec:handshake:traffic}

\begin{figure}
	\centering
	\includegraphics[width=.8\columnwidth]{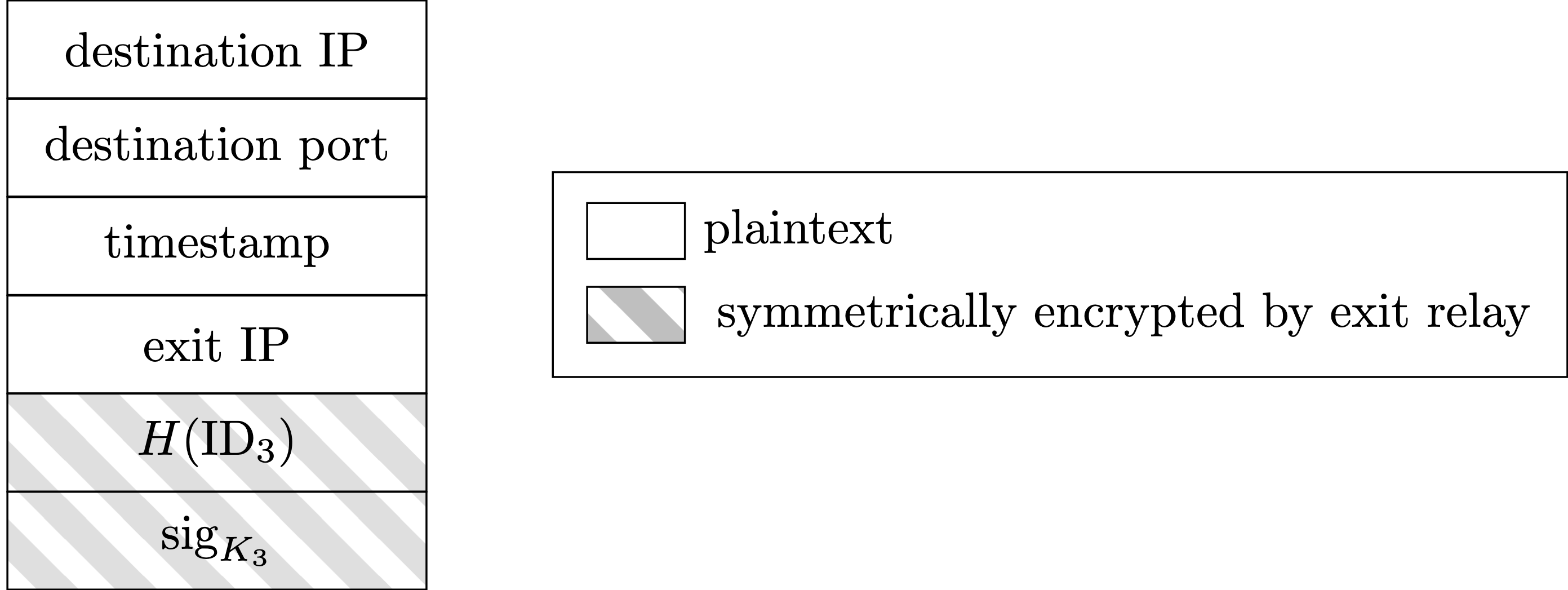}
	\caption{Flow records are searchable by \acp{LEA}. The link to the encrypted identity is however encrypted by the exit relay to prevent the matching of multiple flows to the same circuit.
  Storing a flow ultimately produces \SI{134}{B} of data.}
	\label{fig:connection-info}
\end{figure}

Following an established circuit, the exit relay's outgoing traffic must be indisputably linked to a client.
Here, \name takes a flow-based approach.
Therefore, whenever the client makes the exit relay establish a new flow, it signs this \emph{flow record} and the database key to its encrypted identity $H(\text{ID}_3)$ with the client identification key.
The flow record, illustrated in Figure~\ref{fig:connection-info}, consists of the destination's IP address, the targeted port, the exit relay's IP address, and a timestamp.
Only the client can generate this signature, and the corresponding public key is stored in the encrypted identity $\text{ID}_3$.
In case of a deanonymization request, the flow record, this signature, and the encrypted identity shift the blame toward the client.

In practice, when opening a new connection, the client creates the signature over the flow record and includes it in the first transmitted cell.
The exit relay verifies this signature and then establishes the connection.
Afterward, the client can communicate over this flow without additional overhead through the circuit.
To outsource the storage of this flow record, the exit relay symmetrically encrypts the signature and the digest $H(\text{ID}_3)$.
The resulting data layout is displayed in Figure~\ref{fig:connection-info}.
We made the design choice to symmetrically encrypt this sensitive data and rely on the exit relay to decrypt it in case of a deanonymization request.
Even if the exit relay were offline, it would be traceable through the consensus or IP address.
With another threshold encryption, we could avoid this involvement of the exit relay, but this would reduce oversight.
Most importantly, the active involvement of exit relays ensures the correctness of the statistics and disclosure of deanonymization requests even for a dishonest consortium.

In theory, two different clients could open a connection through the same exit relay in close succession.
In this case, it would be hard to match observed Internet traffic to a specific flow record.
This challenge stems from the fact that the client cannot reliably know the port used by an exit relay; hence, the client cannot include this information in its signature.
In practice, we expect such similar flow records to rarely and mostly happen to widely known and benign websites, \eg news sources.
In the worst case, \acp{LEA} would have to investigate two or more circuits, which could come with higher scrutiny by the consortium deciding on the deanonymization request.

An additional challenge comes from the fact that no single database operator is trusted by everybody.
It would, however, put additional strain on the exit relays if they have to migrate their database entries to multiple deployed databases. 
We offload these synchronization costs to the operators while simultaneously benefiting from the better connectivity among them. 
However, if the operators synchronize the databases themselves, one operator must be sure that they know about all data distributed by the exit relays.
Hence, we implemented a smart contract (deployed on the translucent ledger introduced in Section~\ref{sec:recovery:ledger}) through which exit relays inform about the digest of migrated database entries. 
Each database operator thus learns which data other instances should distribute to them.
All in all, \name thus allows the distributed storage of the necessary data for deanonymizations without overburdening the exit relays or other volunteers with additional computations or costs.

\subsection{A \name Circuit from the Perspective of an Exit Relay}

\begin{figure}
	\centering
	\includegraphics[width=\columnwidth]{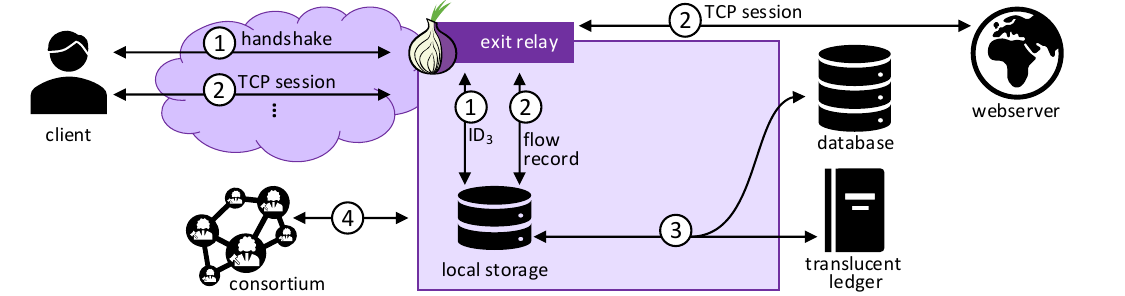}
	\caption{During the establishment of a circuit, a \name exit relay first \stepnumberspace{1}{performs} the \name handshake and stores the encrypted identity of the client. Then, \stepnumberspace{2}{one} or multiple connections are opened by the client and the flow records stored by the exit relay.
  The exit relay \stepnumberspace{3}{periodically} publishes its locally stored data to the database and \stepnumberspace{4}{remains} available for possible collaboration with the consortium during deanonymizations. 
  }
	\label{fig:exit}
\end{figure}

We now present the anonymization procedure of \name from the perspective of an exit relay to summarize the system.
As shown in Figure~\ref{fig:exit}, the exit relay's involvement \stepnumberspace{1}{begins} with the last part of the handshake to establish a new circuit for the client. 
It validates the client identification data, computes the encrypted identity $ID_3$, and assigns the extracted client identification key to the circuit as described in Section~\ref{sec:handshake:exit}.
Then, the exit relay stores $ID_3$ locally.

Next, the client \stepnumberspace{2}{establishes} TCP sessions through the circuit.
While establishing the connection, the client submits a flow record signed by its identification key to the exit relay. 
The exit relay extracts this flow record and verifies it.
Then, the flow record is stored locally by the exit relay, which then establishes the connection.
The client can now send data like in Tor.

Finally, the exit relay \stepnumberspace{3}{regularly}  publishes its locally stored encrypted identities and flow records to the database.
To ensure consistency in the database, this also involves writing the hash of each update to a smart contract on the translucent ledger.
Moreover, the exit relay \stepnumberspace{4}{remains} available for potential deanonymization requests by the consortium.

\section{Deanonymization Process}
\label{sec:deanonymization}

\begin{figure}
	\centering
	\includegraphics[width=.8\columnwidth]{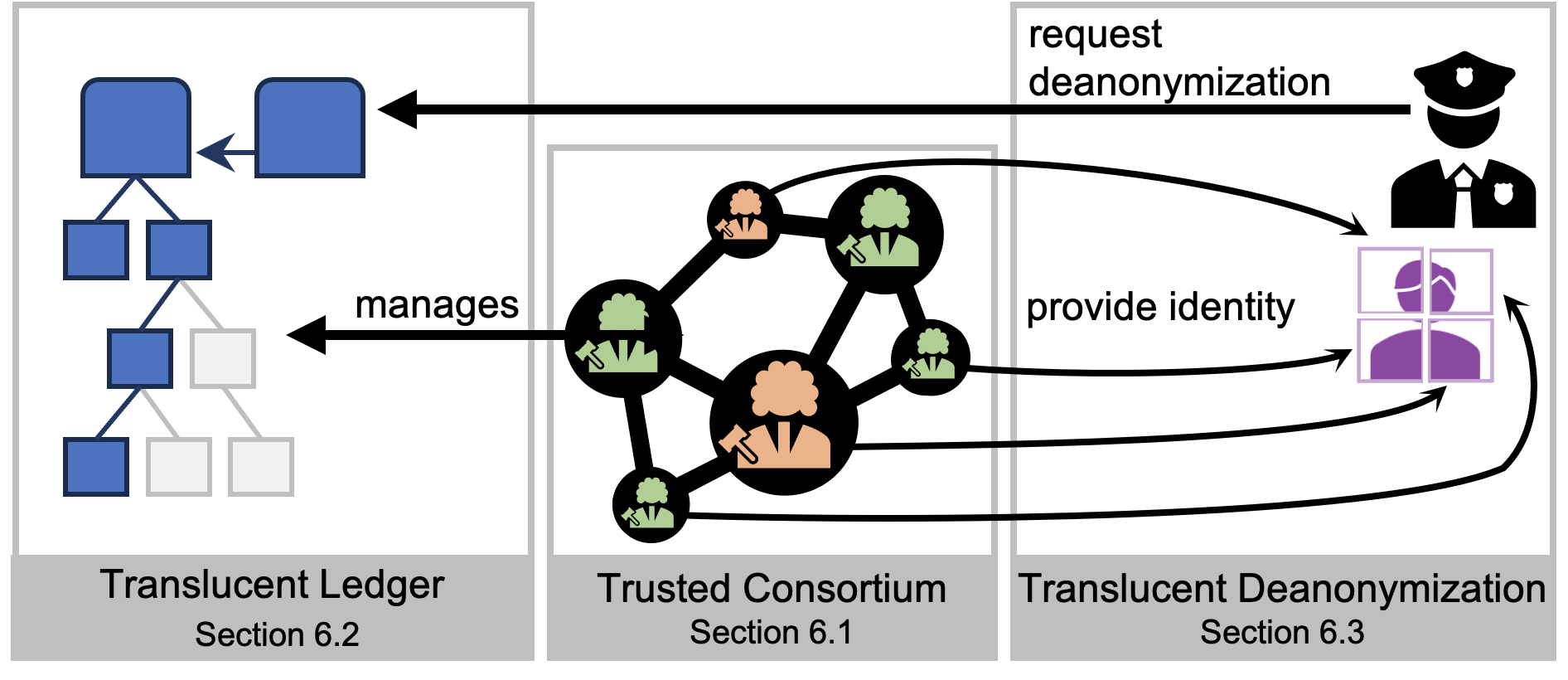}
	\caption{ \name relies on a trusted consortium. During operation, \acp{LEA} can make deanonymization requests on a translucent ledger, which will be voted on by the consortium and, if accepted, collaboratively executed by the consortium.}
	\label{fig:recovery}
\end{figure}

In the following, we look at \name's deanonymization process, which is illustrated by Figure~\ref{fig:recovery}.
\name relies on a consortium of trusted parties to approve and execute deanonymizations.
While this paper focuses on the technical aspects, we nevertheless give some considerations about the potential composition of such a consortium in Section~\ref{sec:recovery:consortium}.
To ensure public oversight, we introduce \emph{translucent ledgers} in Section~\ref{sec:recovery:ledger} to manage this consortium.
The use of a translucent ledger enables the public to see live statistics about attempted and successful deanonymization requests, ensures the delayed information disclosure, and allows the consortium to grant selective data access to prove an approved request to an exit relay.
Afterward, we discuss the deanonymization process in Section~\ref{sec:recovery:process}.
Finally, we see how \name probes relays to preemptively detect rogue relays without compromising anyone's privacy in Section~\ref{sec:recovery:probing}.

\subsection{Assembling a Trusted Consortium}
\label{sec:recovery:consortium}

We identify three roles that should be represented in \name's consortium: \emph{governments}, \emph{privacy advocates}, and \emph{privacy-focused corporations}.
First, we need representatives from the judicial branch of participating governments.
These entities have the knowledge to assess deanonymization requests based on the potential benefit to public safety and their legality.
Secondly, the consortium needs to incorporate parties that are known to strongly advocate for people's privacy.
These parties are essential to generate trust in the system. 
They can block excessive requests or such requests that clearly do not benefit public safety, \eg if the deanonymization of a whistleblower would be requested.
This group could include the operators of Tor's directory authorities and other well-known privacy advocates.
Thirdly, we see a need to include corporations that are known to prevent governments from excessively accessing encrypted data, \eg \emph{Mozilla} and \emph{Apple}.
These are powerful entities relative to typical privacy advocates and thus are better equipped to oppose political pressure.

Besides the type of representatives in the consortium, its size must also be determined, as well as the threshold to approve requests.
This threshold can be lower than the threshold set for the threshold decryption, assuming that non-agreeing entities still cooperate in the decryption once a vote has passed.
Moreover, the interval between the renewing of threshold-encryption keys must be decided on, such that honest consortium members can ensure that no data can be accessed beyond its intended retention period.
Overall, the consortium should be relatively static, and changes should be planned in advance to happen with scheduled key updates, but they also require updating the smart contract.
Considerations regarding the consortium's size and the voting procedure (\eg simple majority, majority within each role, etc.) remain important aspects to \name's deployment but go beyond the scope of this work.

\subsection{Translucent Ledgers}
\label{sec:recovery:ledger} 

We propose the concept of \emph{translucent ledgers} for the coordination among the consortium while ensuring public oversight.
Generally, a translucent ledger is a private blockchain where a new block (deanonymization requests in \name's case) is accepted by the consensus of the consortium members and is cryptographically linked to its predecessor.
For \name, we use an Ethereum~\cite{2014_wood_ethereum} blockchain with the \ac{PoA} consensus mechanism, which requires minimal modifications to be used as the translucent ledger.
What makes a ledger translucent is that the consortium grants access to selected data for specific clients and the public.
The three pillars of translucency are (i) delayed disclosure, (ii) peeks into real-time data, and (iii) real-time statistics.
These three pillars allow keeping ledger data secret as much as necessary while providing guarantees of the correct execution of processes to the public.

\textbf{Delayed Disclosure.}
The first building block of translucent ledgers is the notion of delayed disclosure, \ie the delayed release of all stored information to the public such that they can be verified by everyone.
For \name, this delayed release is based on a fixed delay, as already exists for the declassification of secret documents by democratic governments.
Still, the consortium may unanimously agree on delaying the release of specific data, \eg if a case is still ongoing.
Here, a single honest entity in the consortium suffices to enforce the timely data disclosure, while the blockchain properties ensure data integrity.
Importantly, this step only reveals which deanonymizations have been requested (including a reasoning) and which of those succeeded, while the client's identity is never written to the ledger and thus also not revealed to the public.

\textbf{Peeks into Real-time Data.}
While disclosure to the public is delayed, interactions with the ledger must happen in real time.
Especially exit relays must be able to read the ledger when they are authorized to do so, \ie to verify the approval of a deanonymization request.
Yet, no additional information should leak about the ledger's current state to unauthorized entities.
To enable such fine-grained interactions, the operators of translucent ledgers can generate verifiable claims of the current state of parts of the ledger.
We build our translucent ledger based on the Ethereum blockchain, such that we can use Merkle proofs as proposed in EIP1186~\cite{2018_jentzsch_eip1186} to reveal the current state of each smart contract variable individually by linking it to the public header.
These Merkle proofs link the revealed data to the already distributed and immutable block headers known to the public.
Thus, authorized outsiders are enabled to peek into specific parts of the ledger's state.

\textbf{Real-Time Statistics.}
Delaying information disclosure for a significant time can make a process appear opaque, making users more skeptical of its trustworthiness.
To counteract this possibility, translucent ledgers can distribute real-time statistics about what is being stored and processed on the ledger.
To this end, we use the Bloom filters in Ethereum~\cite{2014_wood_ethereum} that usually are used to let clients quickly search for past events.
Events are introduced by smart contract programmers and are triggered when certain conditions are met during code execution.
Upon being fired, these events write logs to the ledger.
Additionally, the specific event that was triggered and its arguments are inserted into the Bloom filter of the current block header.
Thus, by strategically specifying these events, a smart contract developer can control what real-time statistics are leaked into the block header.

Translucent ledgers are thus a strong building block for the management of exceptional access that is neither opaque nor completely transparent to the public.

\subsection{Translucent Selective Deanonymizations}
\label{sec:recovery:process}

We now discuss how we use a translucent ledger to create a translucent deanonymization process.
After identifying suspicious \name-anonymized traffic, an \ac{LEA} requests a deanonymization from the consortium.
Therefore, the consortium first assesses this request through a translucent voting.
If the request is approved, the consortium members, with the help of the exit relay, jointly remove the layers of threshold encryption from the client's encrypted identity.
The final decryption shares are then only shared with the requesting \ac*{LEA}, such that the identity of the suspect is not even revealed to the consortium.

\subsubsection{Translucent Voting}

We assume that \acp{LEA} monitor the Internet or request logs from websites to identify suspicious traffic that has been anonymized by \name when deanonymization is warranted to uphold public safety. 
As a first step, the \ac{LEA} then searches the database for the flow record (\cf~Figure~\ref{fig:connection-info}) that matches the observed traffic.
For now, consider that this database is fully accessible to \acp{LEA}, such that \acp{LEA} have a list of all communication flows exiting \name.
To further strengthen privacy, this database could be based on searchable encryption~\cite{2006_curtmola_searchable} (and rate-limited through cryptographic puzzles or the database being operated by the consortium) where each exit relay logs to a different index with an encryption key only known to the relay.
This index could still be searched by \acp{LEA} to match concrete flow records, while no information about all traffic leaks.

After identifying the appropriate flow record, the \ac{LEA} can request a deanonymization (and supply evidence) from the consortium by calling the corresponding function of the smart contract deployed on the translucent blockchain.
This deanonymization request contains one suspicious flow record (referenced by its hash) and justification for the consortium to approve the request, which are both written to the translucent ledger.
What would be accepted as valid justification must be agreed upon by the consortium, but it could involve a warrant or the evidence collected by the \ac{LEA}.
Appendix~\ref{app:smart-contract} shows an exemplary smart contract for a consortium with five members and basic majority voting.
This function call triggers an event in the smart contract which reveals that a deanonymization request occurred immediately to the public, while the suspicious flow record and justification will only later be revealed to the public.

Then, each consortium member can call the smart contract to vote in favor of the deanonymization.
If enough positive votes are collected, the request is accepted; this approval triggers an on-chain event visible to all \name users, \ie the public.
The delayed disclosure then ensures that the individual votes of each consortium member will eventually be fully published.
This event also starts the deanonymization procedure by the consortium.
Following a successful vote, however, only the exit relay can decrypt the key in the flow record linking it to the encrypted identity.

\subsubsection{Linking Flows to Encrypted Identities}
\label{sec:exit-relay}

The consortium needs to request the exit relay to link the identified flow record to an encrypted identity.
Therefore, the exit relay needs to be informed about the flow record and convinced that the request was approved and logged by the consortium.
The consortium provides this proof through an EIP1186 Merkle proof~\cite{2018_jentzsch_eip1186}, which proves the deanonymization request (without justification) and voting record to the exit relay by revealing the hash chain that connects the smart contract state variables to the already public blockchain header.
This communication between the consortium and the exit relay is confidential and realized over a TLS channel between one consortium member and the exit relay.

Thus, only the exit relay is informed about this request, as \name would otherwise extensively interfere with ongoing investigations.
After verifying the request, the exit relay is thus sure that the consortium agreed on the necessity of this deanonymization and that this will eventually be disclosed.
Then, the exit relay decrypts the digest $H(\text{ID}_3)$, which is the database key for the encrypted client identity $\text{ID}_3$, as well as the flow record signature by the client identification key. 
The exit relay then transfers this data to the consortium.

The consortium can now retrieve the associated encrypted identity $\text{ID}_3$~(\cf~Figure~\ref{fig:encrypted-identity}) and remove the first threshold-encryption layer to verify the honesty of the exit relay.
For this verification, several requirements must be fulfilled.
First, the data pointed to by the database key has to be retrievable.
Second, the signature of the flow record must be verifiable with the client identification key $K_3$ revealed by the decryption.
This signature ensures that the outer encrypted identity is indeed the origin of the investigated communication,
as only the client knows the corresponding private key and would not sign flow records of unknown connections.
Third, the consortium has to check that the indicated predecessor relay has a known identity $R^{\text{id}}_2$.
Knowing the middle relay's identity means that either a valid certificate is provided or, more commonly, the relay's identity is stored in \name's consensus.
Finally, the consortium ensures that this middle relay did indeed vouch for the correctness of the next layer's encrypted identity by verifying the signature $\text{sig}_{R_2^{\text{id}}}\left(\text{ID}_2||K_3||R_2^{\text{id}}\right)$.

Together, these verification steps convince the consortium that the exit relay received the claimed inner encrypted identity $\text{ID}_2$ from the indicated predecessor relay and that the exit relay acted according to the protocol.
If any of the preceding verifications fail, this would indicate a malicious exit relay and thus likely collusion with the client. 

\subsubsection{Extracting the Traffic's Origin}

After verifying the exit relay's honesty, the consortium proceeds to similarly verify the other relays' honesty, ultimately leading to the identification of the client's identity.
Therefore, the consortium first jointly decrypts the inner encrypted identity $\text{ID}_2$.
Then, the consortium verifies the honesty of the middle relay with the same process as for the exit relay.
Additionally, the consortium verifies that the hop key $K_3$ matches that of the outer encrypted identity layer and that this key is correctly signed by $\text{sig}_{K_2}\left(K_3\right)$.
Again, if any verification fails, the middle relay behaved maliciously, and the consortium reports this as likely collusion with the client to the \ac{LEA}.
Otherwise, the middle relay acted according to the \name protocol.

If a circuit consists of more than three relays, the same procedure can be used to iteratively unveil further intermediate layers of the encrypted identity.
Only the last layer is treated differently.
While the encrypted identities are handled by the consortium, the plaintext origin of the traffic is only communicated to the requesting \ac{LEA}.
Therefore, the consortium members only share the final decryption shares with the \ac{LEA}.
Thus, no member of the consortium learns this identity.
This restriction maximizes privacy, as only the requesting \ac{LEA} learns the client's identity, while the consortium only knows the relays.

The \ac{LEA} then still has to verify the correctness of the decrypted client identity to ensure that the entry relay did not manipulate this information.
Therefore, the \ac{LEA} first verifies that the IP certificate is valid and that a trusted CA issued it.
Additionally, the \ac{LEA} verifies the correctness of the hop key $K_1$, analogously to this verification by the consortium as previously discussed.
To this end, each consortium member shares the hop key $K_2$ claimed in the previous layer with the \ac{LEA} along its decryption share.
If both checks succeed, the revealed identity is indeed that of the user from whom the suspicious traffic originated.
Otherwise, the entry relay attempted to cover up the true identity, and the \ac{LEA} should investigate this.
Through \name's iterative process, the requested identity is thus only revealed to the requesting \ac*{LEA}, while misbehavior (\ie sending no or false data) during the anonymization process can be clearly attributed to one relay.

\subsection{Probing Relay Honesty}
\label{sec:recovery:probing}

Deanonymizations reveal misbehaving relays.
Such misbehavior could occur due to collusion with the client, general malice toward the network, or from genuine errors.
Therefore, the relay's protocol abidance under normal circumstances should be verified without risking individuals' privacy, such that misbehaving nodes can be fixed or removed from the network.
Regular probing by the directory authorities reduces the likelihood that an exit relay's key loss in a critical situation is coincidental.
This feature significantly strengthens our assumption that failures during deanonymization are due to a relay colluding with a client, warranting further investigation.

To generate probe traffic for verifying the honesty of relays, \name relies on cryptographic commitments.
We use Pedersen commitments~\cite{1991_pedersen_commitment}, but other commitment schemes could also be employed.
To create a commitment, a directory authority, \ie the prober, first generates a key later used as the client's identification key $K_3$ during the probe.
Then, the prober hashes this key, the probed exit relay, a traffic destination, and a timeframe during which the probe is active.
We refer to the resulting digest as probe $p$.
To then create a Pedersen commitment to the probe $p$, the prober generates a random value $r$ and computes the commitment $C = pG + rH$ on an elliptic curve with generators $G$ and $H$.

The prober then invokes a smart contract on the translucent ledger to store $C$.
Afterward, the prober establishes a new circuit with the probed relay as the last onion router and sends traffic as specified in the commitment and using the pre-generated client identification key.
Afterward, the prober opens its commitment $C$ and discloses the preimage of $p$ on the ledger.
Then, the prober performs a deanonymization request, using the probe as justification for the deanonymization.
The consortium accepts this request, and the exit relay is asked to initiate the deanonymization process.

At this point, the exit relay does not know that the deanonymization request is only for probe traffic.
It learns that the vote for the request was successful but does not see the reasoning behind it, which is only revealed during the delayed disclosure of the request to the public.
The exit relay thus is expected to cooperate like for ordinary deanonymization.
Afterward, the consortium verifies the signature of the database key $H\left(\text{ID}_3\right)$ with the public key indicated in the commitment.
If this verification is successful, the exit relay passed the probe.
Otherwise, the directory authority no longer accepts this exit relay.
The other directory authorities can adopt this assessment but are encouraged to verify it through independent probes.
Otherwise, a single directory authority would have the power to remove relays at will by claiming they failed a probe.
However, a malicious relay may try to detect such additional probes and behave honestly for them to not be removed from the network.
Hence, once suspicious, a relay should be probed with a higher frequency at random intervals to ensure any misbehavior was indeed a one-off event.
The consortium can also further investigate the encrypted identity to identify non-exit relays that are misbehaving.
Through regular probing, the consortium and the \acp{LEA} can be much more confident that each relay behaves correctly during deanonymization.

In general, there are two concerns that dictate the selection of a sensible probing frequency: strain on the network and detectability.
The exit bandwidth consumption of one probe is dominated by the size of the accessed web page, where a median web page is \SI{2.8}{MB} in size~\cite{http-archive}.
Having each directory authority probe each exit relay once per day would thus result in not even \SI{1}{MB/s} of exit bandwidth consumption.
To put this overhead into perspective, we can look at bandwidth authorities~\cite{2016_alsabah_performance}, which consume orders of magnitude more bandwidth for relay capacity estimations~\cite{2022_darir_mleflow}.
Moreover, to avoid that probes are detected, the respective deanonymization needs to take time to match the manual decision-making within the consortium.
Given that executing multiple probes in parallel likely adds limited value, a probing frequency above one probe per day is unrealistic.

Overall, probing is expected to make up the bulk of all requests.
As the public is informed about each request immediately, all probers are encouraged to reveal that a probe took place immediately after conclusion.
Therefore, they can call a special function in the smart contract that triggers an event to inform the public, such that the public always knows how many requests actually deanonymize a user.

\section{Performance Evaluation}
\label{sec:perf-intro}

By comparing \name's performance to Tor, we now assess the costs of realizing anonymity with translucent selective deanonymization capabilities.

\subsection{Evaluation Setup}
\label{sec:eval-setup}

Most new Tor features are evaluated through simulations.
However, simulation abstracts from Tor's cryptographic processing~\cite{2012_jansen_shadow}, which is precisely the part impacted by \name (and not the behavior of the network).
Thus, we evaluate \name in a local deployment.
Since our changes do not impact the network as a whole, but rather individual relays or circuits, this still gives accurate results on \name's overhead.

We modify Tor version 0.4.1.5 
to implement \name.
As our target host for connection establishments, we run an \emph{apache2} web server.
We additionally host a local CA for IP certificates and a \emph{MongoDB} database to store deanonymization information.
The consortium members' behavior is expressed in Python, and the translucent ledger is a permissioned Ethereum blockchain~\cite{2014_wood_ethereum} modified to realize the translucency properties.
Our server contains two Intel\textsuperscript{\textregistered} Xeon\textsuperscript{\textregistered} Silver 4116 processors (24 CPU cores\,@\,\SI{2.1}{GHz}) and \SI{204}{GB} of RAM.
We run nine relays: one directory authority, three exit relays, and five non-exit relays.
Each process (\eg databases, Tor clients, Tor relays) is pinned to one CPU core to prevent cross-process interference.

\begin{figure}
	
	\centering
	\includegraphics{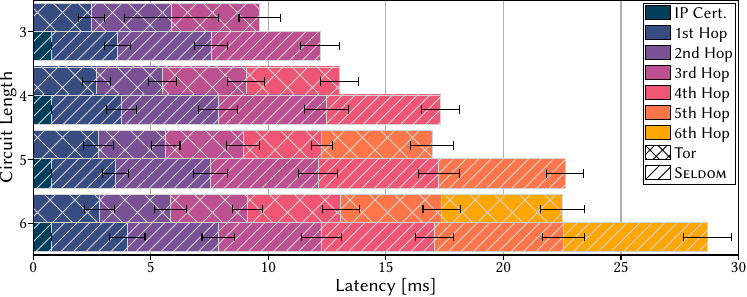}
	\caption{During circuit establishment, \name adds only single digit milliseconds of delay, which is overshadowed by the latency of a global network.
	Error bars show \SI{95}{\%} confidence intervals. The processing overhead of \name is not noticeable to users and thus unlikely to cause an issue.}
	\label{fig:establishment}

\end{figure}

\subsection{Circuit Establishment}
\label{sec:eval:circuit}

First, we measure the overhead of \name during circuit establishment.
All communication happens through the loopback interface, which means that we do not include latency between different entities.
During the evaluation, the client periodically establishes new circuits and logs whenever a circuit is extended by a new hop.
For \name, the client obtains a new IP certificate before each circuit establishment (usually not necessary, as these certificates are reusable and can be obtained in advance).
We establish approx.\ 900 circuits in Tor and \name for each evaluated circuit length.

We compare the latency to extend a circuit by one hop at a time in Figure~\ref*{fig:establishment}.
The x-axis shows the elapsed time, while the different horizontal bars represent circuits of various lengths in \name or Tor, respectively.
The cumulative bars represent the total time to establish a circuit.

We measure a processing overhead to extend the circuit by one hop that increases from \SI{7.4}{\%} for the first hop to \SI{19.6}{\%} for the sixth hop.
This increase arises from the larger encrypted identities of longer circuits.
Overall, we observe a resulting latency increase of \SI{3.2}{ms} to \SI{4.3}{ms}, depending on circuit length.
The receipt of an IP certificate causes an overhead of \SI{0.8}{ms}. 
The remaining overhead stems from signature processing, threshold encryption, and transferring the partially encrypted identity to the next hop.
All these steps require more time the more previous hops exist (as more data has to be processed).
The observed variance in timings results from occasional failures of the path selection algorithm due to the small network size.
In a global overlay network, additional latency in the range of multiple 100 milliseconds and jitter would be added by the network and make this overhead insignificant.
Overall, \name thus does not perceivably impact circuit establishment.

\begin{figure}
	\centering
	\includegraphics[width=.8\columnwidth]{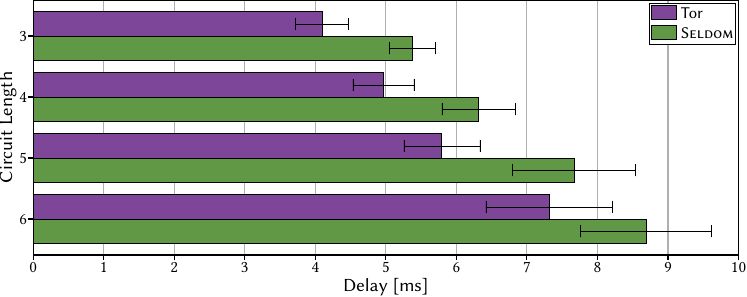}
	\caption{The average latency to retrieve the first byte when connecting to a new host through an established circuit shows barely perceivable overhead by \name.}
	\label{fig:eval:connection}
\end{figure}

\subsection{Connection Establishment}
\label{sec:eval:connection}

Analyzing the impact of \name on connection establishment
is essential, as additional latency in this step directly affects the user experience.
To assess this overhead, we request a web page from the locally hosted web server through \textit{cURL}, which uses \textit{socks5} to communicate with the Tor client.
We measure the time to first byte, \ie the time between issuing the request and receiving the first byte through the established circuit.
We repeat this measurement for different circuit lengths (approx.\ 700 measurements each) and compare a vanilla Tor deployment to \name.
Our results are shown in Figure~\ref{fig:eval:connection}.
We observe that \name takes, on average, \SI{1.5}{ms} longer than Tor to establish a connection and receive the first byte of the response, independent of the circuit length.
This delay stems from the generation and verification of the flow record signature as well as writing this flow record to a local database before the connection is established.
Again, these delays are overshadowed by the latency of the overlay network, which is two orders of magnitude higher.

\subsection{Bandwidth and Storage Requirements}
\label{sec:eval-longitudinal}

While \name barely impacts performance, the additional data that is transferred and stored risks becoming the real bottleneck.
To quantify this overhead, we first measure how much bandwidth exit relays consume to transfer all deanonymization data to a state-sponsored database.
During our evaluation, the client establishes circuits with a length of three and a fixed exit relay.
A fixed number of connections is established to random IP addresses and ports through each circuit before it is discarded.
The exit relay waits for a predefined amount of closed circuits before it migrates the combined local storage for these circuits to the state-sponsored database.
When migrating, \textit{mongodump} is executed on the local \textit{MongoDB} database, and the resulting folder is serialized and compressed using \textit{zlib}.
The resulting archive is sent over a TLS channel to a state-sponsored database.
Simultaneously, its hash is sent to 
the translucent ledger to ensure consistency across multiple governmental databases.

\begin{figure}
	\centering
	\includegraphics[width=.8\columnwidth]{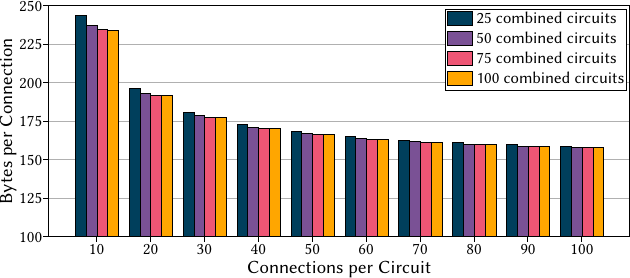}
	\caption{Considering all data that is transmitted when flushing deanonymization data to the database, each connection consumes between 156 and {244}{B} of data. This overhead shrinks the more connections are established per circuit and as more circuit's data is flushed together.}
	\label{fig:eval:bandwidth}
\end{figure}

In Figure~\ref{fig:eval:bandwidth}, we show how bandwidth usage depends on the number of connections established per circuit and how often the local buffer is flushed. 
Flow records are approximately an order of magnitude smaller than encrypted identities, which are only generated for each circuit and not each connection.
Furthermore, the more data is migrated to the database in one go, the more the local buffer can be compressed.
However, the benefit of sending more data at once levels off quickly, indicating that frequent flushing is desirable. 
While using an established circuit for more connections is beneficial, such a change directly influences the susceptibility to deanonymization attacks.
Recent measurements by Mani~\etal estimate the average number of established connections per circuit to be about 20~\cite{2018_mani_tor-measurements}.
An exit relay that migrates its database after 100 circuits have been established and used to establish 20 connections each would thus produce \SI{191.7}{byte} of data per connection.

The measurements of Mani~\etal also allow us to predict the storage requirements of a Tor-sized \name network.
For this prediction, we project the amount of generated deanonymization data based on the activity in the Tor network over time.
According to Mani~\etal, an average of 2.1 billion (95\%-CI:[1.7 billion; 2.5 billion]) exit streams were created every 24 hours in April 2018~\cite{2018_mani_tor-measurements}.
Over the same period, the Tor Project reported an average of \SI{111.55}{Gbit/s} of consumed bandwidth by the Tor network~\cite{tor-metrics}.
For our estimations, we assume a constant relationship between the amount of consumed bandwidth and the amount of exit streams over the last six years.
This is a worst-case assumption, as the size of web pages grew by over \SI{29}{\%}, 
while the number of TCP connections per web page shrunk by about \SI{47}{\%} over the same period~\cite{http-archive}.
Thus, in practice, the additional bandwidth consumed by Tor is, in part, caused by larger web pages that are downloaded over fewer connections.

\begin{figure}
	\centering
	\includegraphics[width=.8\columnwidth]{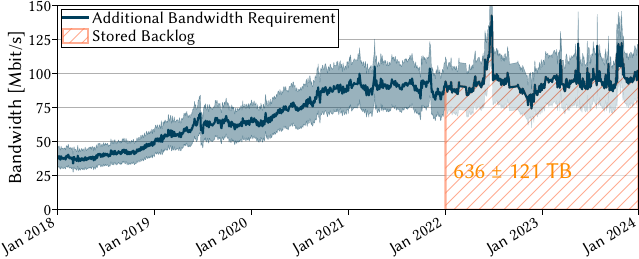}
	\caption{\name increases exit bandwidth by $0.1\%$ to store deanonymization data. The total storage requirement for a retention period of two years amounts to $636 \pm 121$\,TB.}
	\label{fig:eval:storage}
\end{figure}

Figure~\ref{fig:eval:storage} thus quantifies the bandwidth overhead of \name in a Tor-sized network according to those estimations.
Here, the blue line shows the global bandwidth consumed by all exit relays to write data into the state-sponsored database over time with \SI{95}{\%} confidence intervals (blue area).
In January 2020, all exit relays combined would consume \SI{67.7}{Mbit/s} (95\%-CI:[\SI{54.8}{Mbit/s}; \SI{80.6}{Mbit/s}]) to write deanonymization data to the state-sponsored database.
To put these numbers into perspective, this would mean a mere \SI{0.11}{\%} (95\%-CI:[\SI{0.09}{\%}; \SI{0.013}{\%}]) increase in data transmitted by exit relays compared to Tor.
A \name exit relay advertising \SI{50}{Mbit/s} of exit traffic per second would thus insert approx.\ \SI{56}{kB/s} of deanonymization data into the database. 
This estimate also means that this same relay has to, on average, verify 385 \code{ed25519} signatures and compute 18 hybrid threshold encryptions per second, a processing overhead hardly impacting exit relays.
Figure~\ref{fig:eval:storage} also highlights the amount of data that would be retained in the state-sponsored database, assuming a retention period of two years.
The database would thus have to store \SI{636}{TB} (95\%-CI:[\SI{515}{TB}; \SI{757}{TB}]) of deanonymization data, which can easily be provided by government organizations.

\name can thus easily scale to Tor's current size.
If \name were to significantly outgrow Tor, there exist several strategies to improve scalability.
First, websites that are unlikely to be abused for illegal activities (\eg news websites) could be whitelisted, \ie exit relays would not store flow records (and encrypted identities if only such websites are visited).
Secondly, \name could rely on \emph{probabilistic deanonymization}, where only a randomly selected subset of identities is deanonymizable.
Thus, the consortium could retroactively select a fraction of records that are persistently stored through provable randomness.
Thus, bandwidth and storage overhead are unlikely to limit \name's scalability. 

\begin{figure}[t]
	\centering
	\includegraphics[width=.8\columnwidth]{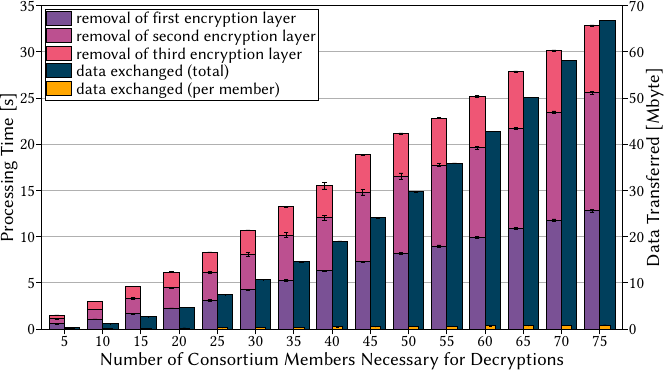}
	\caption{
		Even large consortia deanonymize an identity in under a minute with minimal data exchange. Humans in the loop are expected to cause significantly larger overhead.
	}
	\label{fig:eval:recovery}
\end{figure}

\subsection{Costs of Deanonymizations}
\label{sec:eval-recovery}

Finally, we measure the processing overhead and data transfers during the deanonymization process.
Here, we look at an increasing threshold of required consortium members participating in the deanonymization.
The total size of the consortium does not influence the performance.
For each measurement, we consider a new \name-anonymized circuit and start with the deanonymization request by an \ac{LEA}.
We repeat each measurement 40 times.
We ignore network latency (negligible) and assume instantaneous decision-making by humans in the loop (which would likely take hours to days in real life).

We plot the time required to decrypt the layer-encrypted identities in Figure~\ref{fig:eval:recovery} (stacked bars on the left of each group).
Here, the time for the first decryption layer also includes the time for the flow record.
Potentially surprisingly, the last step in the decryption is notably faster than the previous steps.
This speedup results from decryption shares being collected only by the \ac{LEA} and not all consortium members.
Overall, we observe a linear growth in processing with an increase in consortium members.
Still, even for 75 members, processing on a single CPU core per participant requires only \SI{32.8}{seconds} (95\%-CI:[\SI{32.5}{s}; \SI{33.2}{s}]).
Thus, processing is orders of magnitude faster than human processes and does not constitute a processing bottleneck.

The other relevant performance metric is the amount of exchanged data.
We thus report the total (dark blue bar) and per-consortium member (yellow bar) amount of application layer data sent and received by each party  during deanonymization in Figure~\ref{fig:eval:recovery}.
For the exchanged data per consortium member, we observe linear growth with the number of participants.
As all data is broadcasted to all participants, we observe quadratic growth in the total amount of exchanged data.
Still, the total amount of exchanged data is well below \SI{100}{MB} and thus insignificant.
Overall, we conclude that the deanonymization process is efficient, and manual processes outside \name{}, such as the identification of suspicious traffic and the assessment of requests, are significantly slower.

\section{Security Discussion}
\label{sec:sec-intro}

After having evaluated its performance, we now discuss how \name protects against potential threats.
\name's threat model~(\cf~Section~\ref{sec:threat_model}) discusses the two potential goals of an adversary concerning its deanonymization capabilities:
\begin{inparaenum}[(i)]
    \item evade deanonymizations, possibly by misattributing suspicious traffic to an innocent client, or
    \item conduct unauthorized deanonymizations.
\end{inparaenum}

To analyze how \name thwarts attempts to bypass its deanonymization procedure, we examine the individual actors within the network.
While the database operators and the consortium as a collective are considered trustworthy, relays, CAs, and individual consortium members may act maliciously~(\cf~Section~\ref{sec:trust_assumptions}).
In the following, we analyze how \name technically prevents individuals and colluding entities from evading deanonymizations~(Section~\ref{sec:sec:evade}) or performing unauthorized deanonymizations~(Section~\ref{sec:sec:deanonymize}).
Finally, we discuss additional forms of collusion~(Section~\ref{sec:other-collusion}) that lie outside the scope of \name's direct technical protections.

\subsection{Evading Deanonymizations}
\label{sec:sec:evade}

First, we look at how \name prevents entities from evading deanonymizations.
No relay, consortium member, or client should be able to tamper with deanonymization without being clearly identified as the culprit.
Malicious relays, as well as consortium members, could try to mix deanonymization data across different circuits to mount an attack.
Likewise, clients could route traffic through a self-hosted relay to cover their tracks. 
In the following, we see how the \name protocol protects against such attacks.

\subsubsection{Reusing Encrypted Identities}
A misbehaving relay may try to hinder deanonymizations by reusing the encrypted identity of a previous client, thereby protecting the real client's anonymity and even blaming an innocent client for potentially suspicious traffic.
Here, the attacker's idea is that during deanonymizations, the wrong encrypted identity is decrypted.
Therefore, the relay could include an encrypted identity $\text{ID}_n$ during two different relay extension handshakes.
However, in this case, the hop key $K_{n+1}$ in the encrypted identity would not match the advertised hop key $K_{n+1}$ in the following encrypted identity $\text{ID}_{n+1}$.
This mismatch can be undeniably blamed on the relay during deanonymizations.

Alternatively, the relay could advertise an encrypted identity $\text{ID}_n$, the hop key $K_{n+1}$, relay identity $R^{\text{id}}_n$, and its signatures on two different relay extensions.
In this case, the next relay would, however, never receive a valid signature $sig_{K_{n+1}}(K_{n+2})$ during the next step of the circuit extension.
Hence, the circuit would never be established.
In neither case is it possible for a malicious relay to undetectably reuse an encrypted identity to extend another client's circuit.

\subsubsection{Selectively Misbehaving Self-Hosted Relay}
Similarly, malicious clients could attempt to cover up their traces by hosting their own relay.
This relay would behave honestly for all other clients but incorrectly process the encrypted identity when a client creates a circuit through its own relay.
In this case, the deanonymization process cannot identify the client.
However, the deanonymization process ensures that the self-hosted relay is identified as misbehaving.
Hence, the relay operator and consequently the client are identified as suspicious and will be further investigated by the \ac{LEA}.

\subsection{Deanonymization Attacks}
\label{sec:sec:deanonymize}

While some adversaries may want to prevent deanonymizations through \name, others may aim to perform unauthorized deanonymizations of clients.
In the following, we look at different attacker strategies, ranging from classical traffic analysis to collusion attacks, and how \name defends against these strategies.

\subsubsection{Traffic Analysis Attacks}
\label{sec:sec-traffic-analysis}

Traffic analysis is a common attack to deanonymize users of anonymity networks~\cite{2005_murdoch_traffic-analysis, panchenko2011website}.
While Tor does not explicitly protect against global adversaries~\cite{2004_dingledine_tor}, \ie an entity that intercepts all traffic, analyzing traffic between two or multiple nodes is often possible for adversaries.
For our discussion, we focus particularly on the additional information that may be leaked about the client in \name compared to Tor.
An observer intercepting the outgoing traffic of a client not using a bridge relay learns two things: the hop key $K_1$ of the circuit and the client's intention to use \name.
The hop key $K_1$ does not help in deanonymizing a circuit, and the intention to use the network is also leaked in Tor.

As the relays communicate through TLS channels, even during circuit establishment, no additional information is leaked besides the amount of transmitted data. 
However, this information is unlikely to give an adversary an advantage in detecting circuit extensions, as, usually, many circuits are multiplexed over a single link.
In the end, such attacks can also be executed against traffic patterns of different websites~\cite{panchenko2011website,panchenko2016website,hayes2016kfingerprinting}, which is possible for both Tor and \name.
All traffic leaving exit relays is either anonymized client traffic or encrypted deanonymization information.
Neither leaks any additional information.

\subsubsection{Mismatching Encrypted Identities}
A malicious consortium member could potentially use partial deanonymizations to leak information about users.
Therefore, the malicious member could set up a circuit that includes the threshold-encrypted identity of a targeted user.
The consortium could be tricked to decrypt this identity, \eg by marking the circuit as probe traffic.

However, such an attack would be detected after the consortium removes one layer of threshold encryption due to an invalid signature. 
The corresponding relay would be flagged as suspicious, but the malicious entity, as part of the consortium, would learn the identity of the preceding relay in the targeted circuit.
The attacker could repeat this step and eventually learn the client's identity.
However, the consortium would notice the second iteration of this attack as the plaintexts are from different layers of the same circuit and thus clearly identify an attack from within the consortium. 
Thus, the client's identity remains secure, as long as the attacker does not operate the entry relay of the targeted user.
Additionally, the rest of the consortium and eventually also the public would learn about this one-off attack and could abandon the network or hold the consortium member accountable.

\subsubsection{Collusion among Relays}
\label{sec:sec-anonym}

\name enables the exit relay to learn the encrypted client identity (\cf~Figure~\ref{fig:encrypted-identity}) and indisputably link connections to it while preserving client anonymity (\cf~Section~\ref{sec:authentication}). 
To show that this claim holds, we analyze which relays have access to which data and derive whether any relay has unintended access to client-identifying data.
Here, we consider information that can be correlated within one circuit and across different circuits.

During circuit establishment, the entry relay learns the client's IP certificate, the hop keys $K_1$ and $K_2$, as well as the signature created with $K_1$.
The certificate (and consequently the reuse of $K_1$ across multiple circuit establishments) contains no new information about the client, as the entry relay knows the client's IP address and the current time.

The hop key $K_2$ and, consequently, the signature created with $K_1$, are unique per circuit and derived from a secure source of randomness.
Thus, they do not allow any relay to gain information about clients from other circuits.
The $n$-th middle relay knows the hop keys $K_n$ and $K_{n+1}$, the identity of the previous relay, and signatures by the previous relay and $K_n$.
All other information is threshold-encrypted by the $(n-1)$-th relay.
This nested encryption ensures that no earlier relay can match any information to deanonymize the client.
In Tor, intermediary relays also know their direct neighbors.
As $K_n$ is only shared with the preceding relay and $K_{n+1}$ only with the following relay, non-connected relays cannot notice that they are part of the same circuit. 
Crucially, the signature as the only link between these two keys is not shared with any other relay.

Finally, all signatures are derived from different keys over different data; thus, no information is leaked.
The exit relay also learns about the traffic sent through the circuit (as in Tor) and stores this information in encrypted form.
Additionally, matching different flow records to the same circuit is also not possible, as the circuit-identifying information is encrypted.
Thus, as we prevent information leakage for each data field known to the different relays, the client remains anonymous. 

\subsection{Other Collusion Scenarios}
\label{sec:other-collusion}

Beyond the threats that \name explicitly defends against, additional collusion scenarios remain outside the scope of direct technical protection.
Nonetheless, collusion between a CA and an entry relay as well as collusion among consortium members, is unlikely to materially compromise user anonymity.

\subsubsection{CA and Entry Relay Collusion}
A potential protocol-based attack arises from the possible collusion between a CA and an entry relay.
Here, the malicious entities could create a fictional client with any IP address that is certified by the CA.
If the circuit is then deanonymized for some reason, \eg suspicious activities by the fictional client, the deanonymization process reveals the fake IP address.
In this case, a powerful adversary could thus use \name to incriminate any person browsing the Internet.
However, such incrimination would also be possible with other means by such adversaries.
Most importantly, \name only removes anonymity and gives \acp{LEA} hints for further investigations.
IP addresses, allegedly used for illegal activities with \name or on other platforms, need to be supported by further evidence, as has been concluded by several courts~\cite{ipruling1,ipruling2,ipruling3}. 
Thus, such collusion by powerful entities does not yield them capabilities they would not have without the existence of \name.

\subsubsection{Collusion Within the Consortium}
\label{sec:collusion-consortium}

To ensure that the risk of being covertly deanonymized in \name through collusion of a majority of consortium members and an exit relay is comparable to Tor, the consortium must be assembled accordingly. 
Here, it is important to recall that Tor does not fully protect the anonymity of its users but only decreases the chances that an attacker can learn a client's identity~\cite{tor-blog}.
One attack to deanonymize a client in Tor is through the collusion of the three relays that make up a client's circuit.
Thus, \name's deanonymizations should be more resilient to collusions than collusion of three randomly selected relays.
As the exit relay is involved in the deanonymization attack in Tor and \name, we can remove it from the equation.

Hence, we compare the risk of collusion in \name's consortium to the risk of collusion of two randomly selected relays.
There are multiple reasons why \name's consortium is significantly less likely to collude than these two relays:
while anyone can deploy massive numbers of relays, the identities of the consortium members are publicly known.
Specifically, it would be necessary to convince a significant fraction of privacy advocates, \eg some of the directory authorities, and foreign governments to partake in such an attack.
Even attempting this could result in significant repercussions, which makes a successful collusion attack among the consortium members highly unlikely.
We also designed \name in such a way that a higher number of members must participate in a decryption than have to accept a deanonymization vote, \ie some disagreeing members have to accept the consensus and cooperate, to further minimize the risk of collusion.

\section{Ethical and Technical Concerns}%
\label{sec:design:faq}

The safety-privacy dilemma remains particularly divisive, with \acp{LEA} and privacy advocates maintaining seemingly fundamentally incompatible positions regarding a problem that requires some sort of compromise~\cite{2024_twenning_ach}.
Fostering the understanding of technical means for enabling exceptional access without the dooming vision of privacy-invasive legislation or hidden backdoors for \acp{LEA}, we address frequent concerns regarding \name's design in the following.

\textbf{Q1 -- Do we advertise for \name to replace Tor?}
No.
\name relies on translucency and public oversight, which means that observed misbehavior by \name's consortium must be met with revocations of these privileges, \ie a migration of the user base back to Tor.
In the far future, \name might attract a significantly larger user base than Tor and thus offer better anonymity even when accounting for selective deanonymization capabilities, especially if users start abandoning Tor.
For \name's security concept to work, there must, however, always exist a quick and easy way to migrate back to an anonymity network without deanonymization capabilities, even if temporarily deserted.
This could be achieved with a browser that lets the users easily, and potentially collaboratively, toggle whether to use Tor or \name.

\textbf{Q2 -- Can citizens trust the consortium?}
The underlying assumption for \name is that a majority of consortium members are trustworthy and value privacy by default, except for extraordinary situations that warrant deanonymization.
Those members therefore have little incentive to partake in collusions, especially considering the high risk of such attempts being revealed to the public through the several translucency properties of \name.
Further, \name protects against individual malicious consortium members.
Namely, the majority of honest consortium members will faithfully delete their secret shares after encrypted identities reach the end of the mandated retention period or in extreme events, such as an imminent government takeover.
This way, the consortium renders older encrypted identities unusable for malicious members even in the case of an anticipated dishonest majority gaining control over \name.
However, maintaining \emph{transparency about the goals and non-goals} of \name remains important to avoid creating false expectations about \name's protection.

\textbf{Q3 -- Can individual relays misbehave?}
Once a relay misbehaves and does not comply with deanonymization requests, they will become suspicious of the requesting \ac{LEA}.
Hence, they would reveal themselves as targets for further investigation.
To reduce the number of misbehaving relays, \name can further use active probing (\cf~Section~\ref{sec:recovery:probing}) and remove any relays that misbehave.

\textbf{Q4 -- Can investigations fail then?}
In exceptional cases, \acp{LEA} will not be able to find the true root of harmful traffic using \name alone.
For instance, misbehaving relays may cover up specific bad actors, or bad actors may establish circuits via compromised devices.
However, all these cases lead \acp{LEA} to a distinct location worth of further forensic investigation.
Hence, \name will not always guarantee an immediate investigation success, but it provides crucial pointers to narrow down additional steps.

\textbf{Q5 -- Can \name support hidden services?}
Technically, \name can integrate hidden services.
Each introduction and rendezvous point would be associated with an encrypted identity, which could be deanonymized if deemed necessary.
The consortium must, however, consider the increased privacy implications of agreeing to such a request.
If a \name-like system were to be deployed, it should, however, be carefully considered whether hidden services should be supported or not.

\textbf{Q6 -- Could \name interfere with Tor's service?}
Initially, \name targets the population of privacy-conscious users who currently shy away from using Tor due to having negative associations by tying Tor to cyber-criminality.
Only if \name is capable of establishing a reputation as the infrastructure of choice for privacy-seeking, law-abiding citizens, a potential user migration from Tor to \name could take place.
As a positive effect, this would actively harm the achievable anonymity for cybercriminals, as they cannot hide among a mass of benign traffic.
On the other hand, it becomes harder to abandon \name if a deserted Tor network offers little anonymity.
To ensure that such a situation does not encourage extensive surveillance by the consortium, it must be easy for the user base to switch back to Tor at once, \eg by offering a Tor and \name-compatible client that, by default, switches to Tor if \name is collectively deemed unsafe.

\section{Conclusion}
With this paper, we investigate the potential of selective deanonymizations for anonymity networks.
We find that \name's design barely impacts performance in comparison to Tor, while data overhead remains manageable.
On the other hand, \name relies on a translucent trusted consortium and potentially encourages the migration of users from Tor, which would weaken the anonymity provided by Tor.
With this paper we thus contribute to understanding the advantages and disadvantages of selective deanonymizations in anonymity networks like \name.
These insights should encourage future research to mitigate drawbacks, and, most importantly, they should foster more open, objective, and technical discussions in inevitable future political debates about exceptional access.

\bibliographystyle{ACM-Reference-Format}
\bibliography{paper}

@inproceedings{2018_mani_tor-measurements,
	title={Understanding Tor Usage with Privacy-Preserving Measurement},
	author={Mani, Akshaya and Wilson-Brown, T and Jansen, Rob and Johnson, Aaron and Sherr, Micah},
	booktitle={Proceedings of the Internet Measurement Conference (IMC '18)},
	year={2018},
	note= {{DOI: \href{https://doi.org/10.1145/3278532.3278549}{10.1145/3278532.3278549}}}
}

@inproceedings{1996_blaze_oblivious_key_escrow,
	title={Oblivious Key Escrow},
	author={Blaze, Matt},
	booktitle={International Workshop on Information Hiding},
	year={1996},
	note= {{DOI: \href{https://doi.org/10.1007/3-540-61996-8\_50}{10.1007/3-540-61996-8\_50}}}
}

@article{1999_bellare_translucent_crypto,
	title={{Translucent Cryptography—An Alternative to Key Escrow, and Its Implementation via Fractional Oblivious Transfer}},
	author={Bellare, Mihir and Rivest, Ronald L},
	journal={Journal of cryptology},
	volume={12},
	number={2},
	year={1999},
	note= {{DOI: \href{https://doi.org/10.1007/PL00003819}{10.1007/PL00003819}}}
}

@inproceedings{1997_bellare_vpke,
	title={{Verifiable Partial Key Escrow}},
	author={Bellare, Mihir and Goldwasser, Shafi},
	booktitle={Proceedings of the 4th ACM conference on Computer and communications security (CCS)},
	year={1997},
	note= {{DOI: \href{https://doi.org/10.1145/266420.266439}{10.1145/266420.266439}}}
}

@inproceedings{2018_wright_crumplezone,
	title={{Crypto Crumple Zones: Enabling Limited Access Without Mass Surveillance}},
	author={Wright, Charles and Varia, Mayank},
	booktitle={IEEE European Symposium on Security and Privacy (EuroS\&P '18)},
	year={2018},
	note= {{DOI: \href{https://doi.org/10.1109/EuroSP.2018.00028}{10.1109/EuroSP.2018.00028}}}
}

@article{1996_denning_taxonomy_pke,
	title={{A Taxonomy for Key Escrow Encryption Systems}},
	author={Denning, Dorothy E and Branstad, Dennis K},
	journal={Communications of the ACM},
	volume={39},
	number={3},
	year={1996},
	note= {{DOI: \href{https://doi.org/10.1145/227234.227239}{10.1145/227234.227239}}}
	
}

@article{2015_abelson_keys_under_doormats2,
	title={{Keys under doormats: mandating insecurity by requiring government access to all data and communications}},
	author={Abelson, Harold and Anderson, Ross and Bellovin, Steven M and Benaloh, Josh and Blaze, Matt and Diffie, Whitfield and Gilmore, John and Green, Matthew and Landau, Susan and Neumann, Peter G and others},
	journal={Journal of Cybersecurity},
	volume={1},
	number={1},
	year={2015},
	note= {{DOI: \href{https://doi.org/10.1093/cybsec/tyv009}{10.1093/cybsec/tyv009}}}
}

@inproceedings{2004_dingledine_tor,
	title={{Tor: The Second-Generation Onion Router}},
	author={Dingledine, Roger and Mathewson, Nick and Syverson, Paul},
	year={2004},
	booktitle = {13th USENIX Security Symposium (USENIX Sec'04)},
	note = {{DOI: \href{https://dl.acm.org/doi/10.5555/1251375.1251396}{10.5555/1251375.1251396}}}
}

@inproceedings{2018_savage_selfescrow,
	title={{Lawful Device Access without Mass Surveillance Risk: A Technical Design Discussion}},
	author={Savage, Stefan},
	booktitle={Proceedings of the 25th ACM SIGSAC Conference on Computer and Communications Security (CCS '18)},
	year={2018},
	note = {{DOI: \href{https://doi.org/10.1145/3243734.3243758}{10.1145/3243734.3243758}}}
}

@inproceedings{2000_shoup_threshold-rsa,
	title={{Practical Threshold Signatures}},
	author={Shoup, Victor},
	booktitle={International Conference on the Theory and Applications of Cryptographic Techniques (Eurocrypt '00)},
	year={2000},
}

@inproceedings{2018_dunna_torblock,
	title={{Analyzing China's Blocking of Unpublished Tor Bridges}},
	author={Dunna, Arun and O'Brien, Ciar{\'a}n and Gill, Phillipa},
	booktitle={8th {USENIX} Workshop on Free and Open Communications on the Internet (FOCI '18)},
	year={2018},
	note = {\href{https://www.usenix.org/system/files/conference/foci18/foci18-paper-dunna.pdf}{https://www.usenix.org/system/files/conference/foci18/ foci18-paper-dunna.pdf}}
}

@misc{tor-metrics,
	author = {{The Tor Project}},
	title = {{Tor | Metrics}},
	year = {2024},
	note = {Last accessed: February 8, 2024. \href{{https://metrics.torproject.org/research.html}}{{https://metrics.torproject.org/research.html}}},
}

@misc{http-archive,
	author = {{http archive}},
	year = {2024},
	title = {{State of the Web}},
	note = {\href{https://httparchive.org/reports/state-of-the-web}{https://httparchive.org/reports/state-of-the-web}. Last accessed: June 28, 2025},
}

@article{2014_wood_ethereum,
	title={{Ethereum: A Secure Decentralised Generalised Transaction Ledger}},
	author={Wood, Gavin and others},
	journal = {Ethereum Project Yellow Paper},
	year={2019},
	note = {\href{https://ethereum.github.io/yellowpaper/paper.pdf}{https://ethereum.github.io/yellowpaper/paper.pdf}}

}

@misc{trump-whistleblowing,
	author = {David Smith},
	year = {2019},
	title = {{Trump condemned for tweets pointing to name of Ukraine whistleblower}},
	note = {\href{https://www.theguardian.com/us-news/2019/dec/27/trump-ukraine-whistleblower-president}{https://www.theguardian.com/us-news/2019/dec/27/trump-ukraine-whistleblower-president}},
	note = {Last accessed: February 8, 2024}
}

@inproceedings{2005_murdoch_traffic-analysis,
	title={Low-cost traffic analysis of Tor},
	author={Murdoch, Steven J and Danezis, George},
	booktitle={IEEE Symposium on Security and Privacy (Oakland)},
	year={2005},
	note= {{DOI: \href{https://doi.org/10.1109/SP.2005.12}{10.1109/SP.2005.12}}}
}

@inproceedings{panchenko2011website,
	title={Website fingerprinting in onion routing based anonymization networks},
	author={Panchenko, Andriy and Niessen, Lukas and Zinnen, Andreas and Engel, Thomas},
	booktitle={Proceedings of the 10th annual ACM workshop on Privacy in the electronic society (WPES)},
	year={2011},
	note = {{DOI: \href{https://doi.org/10.1145/2046556.2046570}{10.1145/2046556.2046570}}}
}

@inproceedings{panchenko2016website,
author = {Panchenko, Andriy and Lanze, Fabian and Zinnen, Andreas and Henze, Martin and Pennekamp, Jan and Wehrle, Klaus and Engel, Thomas},
title = {{Website Fingerprinting at Internet Scale}},
booktitle = {Proceedings of the 23rd Annual Network and Distributed System Security Symposium (NDSS)},
year = {2016},
note = {{DOI: \href{https://doi.org/10.14722/ndss.2016.23477}{10.14722/ndss.2016.23477}}}
}

@inproceedings{juarez2014critical,
  title={A critical evaluation of website fingerprinting attacks},
  author={Juarez, Marc and Afroz, Sadia and Acar, Gunes and Diaz, Claudia and Greenstadt, Rachel},
  booktitle={Proceedings of the 2014 ACM SIGSAC conference on computer and communications security},
  year={2014}
}

@inproceedings{cherubin2022online,
  title={Online website fingerprinting: Evaluating website fingerprinting attacks on tor in the real world},
  author={Cherubin, Giovanni and Jansen, Rob and Troncoso, Carmela},
  booktitle={31st USENIX Security Symposium (USENIX Sec' 22)},
  year={2022}
}

@inproceedings{wang2014effective,
  title={Effective attacks and provable defenses for website fingerprinting},
  author={Wang, Tao and Cai, Xiang and Nithyanand, Rishab and Johnson, Rob and Goldberg, Ian},
  booktitle={23rd USENIX Security Symposium (USENIX Sec' 14)},
  year={2014}
}

@inproceedings{hayes2016kfingerprinting,
  title={k-fingerprinting: A robust scalable website fingerprinting technique},
  author={Hayes, Jamie and Danezis, George},
  booktitle={25th USENIX Security Symposium (USENIX Sec '16)},
  year={2016},
  note = {{DOI: \href{https://dl.acm.org/doi/10.5555/3241094.3241186}{10.5555/3241094.3241186}}}
}

@inproceedings{dyer2012peek,
  title={Peek-a-boo, i still see you: Why efficient traffic analysis countermeasures fail},
  author={Dyer, Kevin P and Coull, Scott E and Ristenpart, Thomas and Shrimpton, Thomas},
  booktitle={2012 IEEE Symposium on Security and Privacy},
  year={2012},
}

@inproceedings{cai2012touching,
  title={Touching from a distance: Website fingerprinting attacks and defenses},
  author={Cai, Xiang and Zhang, Xin Cheng and Joshi, Brijesh and Johnson, Rob},
  booktitle={Proceedings of the 2012 ACM conference on Computer and communications security},
  year={2012}
}

@article{oh2021gandalf,
  title={{GANDaLF: GAN for data-limited fingerprinting}},
  author={Oh, Se Eun and Mathews, Nate and Rahman, Mohammad Saidur and Wright, Matthew and Hopper, Nicholas},
  journal={Proceedings on Privacy Enhancing Technologies},
  volume={2021},
  number={2},
  year={2021}
}

@inproceedings{rimmer2018automated,
  title={Automated Website Fingerprinting Through Deep Learning},
  author={Rimmer, Vera and Preuveneers, Davy and Juarez, Marc and Van Goethem, Tom and Joosen, Wouter},
  booktitle={25th Annual Network and Distributed System Security Symposium},
  year={2018},
}

@inproceedings{sirinam2019triplet,
  title={Triplet fingerprinting: More practical and portable website fingerprinting with n-shot learning},
  author={Sirinam, Payap and Mathews, Nate and Rahman, Mohammad Saidur and Wright, Matthew},
  booktitle={Proceedings of the 2019 ACM SIGSAC Conference on Computer and Communications Security (CCS '19)},
  year={2019}
}

@inproceedings{shen2023subverting,
  title={Subverting website fingerprinting defenses with robust traffic representation},
  author={Shen, Meng and Ji, Kexin and Gao, Zhenbo and Li, Qi and Zhu, Liehuang and Xu, Ke},
  booktitle={32nd USENIX Security Symposium (USENIX Sec '23)},
  year={2023}
}

@article{rahman2020tik,
  title={Tik-Tok: The Utility of Packet Timing in Website Fingerprinting Attacks},
  author={Rahman, Mohammad Saidur and Sirinam, Payap and Mathews, Nate and Gangadhara, Kantha Girish and Wright, Matthew},
  journal={Proceedings on Privacy Enhancing Technologies},
  volume={2020},
  number={3},
  year={2020}
}

@article{bhat2019var,
  title={Var-CNN: A Data-Efficient Website Fingerprinting Attack Based on Deep Learning},
  author={Bhat, Sanjit and Lu, David and Kwon, Albert and Devadas, Srinivas},
  journal={Proceedings on Privacy Enhancing Technologies},
  volume={4},
  year={2019}
}

@misc{tor-users,
	author = {{The Tor Project, Inc.}},
	title = {{Who uses Tor?}},
	year = {2024},
	url = {{https://www.torproject.org/about/torusers.html.en}},
	note = {Last accessed: June 5, 2024},	
}

@misc{tor-blog,
	title = {{Traffic correlation using netflows}},
	year = {2014},
	author = {Arma},
	publisher = {Tor Blog},
	url={{https://blog.torproject.org/traffic-correlation-using-netflows?page=1}},
	note = {Last accessed: May 17, 2020. \href{https://blog.torproject.org/traffic-correlation-using-netflows}{https://blog.torproject.org/traffic-correlation-using-netflows}},
}

@techreport{1981_rabin_oblivious-transfer,
	title={How To Exchange Secrets with Oblivious Transfer.},
	author={Rabin, Michael O},
	year={1981},
	note = {\href{https://eprint.iacr.org/2005/187}{https://eprint.iacr.org/2005/187}}
}

@INPROCEEDINGS{2006_dingledine_anonymity-loves-company,
	author = {Roger Dingledine and Nick Mathewson},
	title = {{Anonymity Loves Company: Usability and the Network Effect}},
	booktitle = {In Proceedings of the Fifth Workshop on the Economics of Information Security (WEIS)},
	year = {2006}
}

@inproceedings{2014_backes_backref,
	title={{BackRef: Accountability in Anonymous Communication Networks}},
	author={Backes, Michael and Clark, Jeremy and Kate, Aniket and Simeonovski, Milivoj and Druschel, Peter},
	booktitle={International Conference on Applied Cryptography and Network Security (ACNS)},
	year={2014},
	note = {{DOI: \href{https://doi.org/10.1007/978-3-319-07536-5\_23}{10.1007/978-3-319-07536-5\_23}}}
}

@misc{2018_jentzsch_eip1186,
	author = {Jentzsch, Simon and Jentzsch, Christoph},
	year = {2018},
	title = {{EIP-1186: RPC-Method to get Merkle Proofs}},
	note = {\href{https://github.com/ethereum/EIPs/issues/1186}{https://github.com/ethereum/EIPs/issues/1186}},
	note = {Last accessed: February 8, 2024},
}

@misc{ipruling1,
	title={{ Media CAT Limited v. Adams \& Ors }},
	note={ (EWPCC). \url{http://www.bailii.org/ew/cases/EWPCC/2011/6.html}},
	year         = {2011},
}

@misc{ipruling2,
	title = {{K-Beech, Inc. v. John Does}},
	note={2:11-cv-03995 (E.D.N.Y.). \url{https://cite.case.law/frd/296/80/}},
	year         = {2012},
}

@misc{ipruling3,
	title = {{Cobbler Nev., LLC v. Gonzales}},
	note={17-35041 (D. Or.). \url{https://law.justia.com/cases/federal/appellate-courts/ca9/17-35041/17-35041-2018-08-27.html}},
	year         = {2016},
}

@inproceedings{2012_jansen_shadow,
	title = {Shadow: Running Tor in a Box for Accurate and Efficient Experimentation},
	author = {Rob Jansen and Nicholas Hopper},
	booktitle = {Proceedings of the 19th Symposium on Network and Distributed System Security (NDSS)},
	year = {2012},

}

@inproceedings{2021_kulshrestha_identifying,
  title={{Identifying Harmful Media in End-to-End Encrypted Communication: Efficient Private Membership Computation}},
  author={Kulshrestha, Anunay and Mayer, Jonathan},
  booktitle={30th USENIX Security Symposium (USENIX Sec '21)},
  year={2021},
  note = {\href{https://www.usenix.org/system/files/sec21-kulshrestha.pdf}{https://www.usenix.org/system/files/sec21-kulshrestha.pdf}}
}

@article{2024_nurmi_investigating,
  title={{Investigating child sexual abuse material availability, searches, and users on the anonymous Tor network for a public health intervention strategy}},
  author={Nurmi, Juha and Paju, Arttu and Brumley, Billy Bob and Insoll, Tegan and Ovaska, Anna K and Soloveva, Valeriia and Vaaranen-Valkonen, Nina and Aaltonen, Mikko and Arroyo, David},
  journal={Scientific Reports},
  volume={14},
  number={1},
  pages={7849},
  year={2024},
  publisher={Nature Publishing Group UK London}
}

@inproceedings{2018_winter_tor,
  title={How do tor users interact with onion services?},
  author={Winter, Philipp and Edmundson, Anne and Roberts, Laura M and Dutkowska-{\.Z}uk, Agnieszka and Chetty, Marshini and Feamster, Nick},
  booktitle={27th USENIX Security Symposium (USENIX Security '18)},
  pages={411--428},
  year={2018}
}

@article{2015_rainie_americans,
  title={{Americans' privacy strategies post-Snowden}},
  author={Rainie, Lee and Madden, Mary},
  year={2015},
  publisher={Pew Research Center},
  note={Last accessed: June 1, 2024. DOI: \href{https://policycommons.net/artifacts/619246/americans-privacy-strategies-post-snowden/1600331/}{20.500.12592/jwvf83}},

}

@inproceedings{2017_gallagher_new,
  title={{New Me: Understanding Expert and Non-Expert Perceptions and Usage of the Tor Anonymity Network}},
  author={Gallagher, Kevin and Patil, Sameer and Memon, Nasir},
  booktitle={Thirteenth Symposium on Usable Privacy and Security (SOUPS '17)},
  pages={385--398},
  year={2017}
}

@article{2023_hiramoto_illicit,
  title={Are illicit drugs a driving force for cryptomarket leadership?},
  author={Hiramoto, Naoki and Tsuchiya, Yoichi},
  journal={Journal of Drug Issues},
  volume={53},
  number={3},
  year={2023},
}

@misc{apple_vs_fbi,
	author = {Sam Thielman},
	year = {2016},
	title = {{Apple v the FBI: what's the beef, how did we get here and what's at stake?}},
	url = {{https://www.theguardian.com/technology/2016/feb/20/apple-fbi-iphone-explainer-san-bernardino}},
	note = {Last accessed: June 5, 2024},
}

@inproceedings{2024_twenning_ach,
author = {Twenning, Leon and Baier, Harald},
title = {{Towards arbitrating in a dispute - on responsible usage of client-side perceptual hashing against illegal content distribution}},
year = {2024},
booktitle = {European Interdisciplinary Cybersecurity Conference (EICC '24)},
}

@inproceedings{2021_green_abuse,
  title={Abuse resistant law enforcement access systems},
  author={Green, Matthew and Kaptchuk, Gabriel and Van Laer, Gijs},
  booktitle={Annual International Conference on the Theory and Applications of Cryptographic Techniques (Eurocrypt '21)},
  year={2021},
  note={ {DOI: \href{https://doi.org/10.1007/978-3-030-77883-5_19}{10.1007/978-3-030-77883-5\_19}} }
}

@inproceedings{2023_bartusek_end,
  title={End-to-end secure messaging with traceability only for illegal content},
  author={Bartusek, James and Garg, Sanjam and Jain, Abhishek and Policharla, Guru-Vamsi},
  booktitle={Annual International Conference on the Theory and Applications of Cryptographic Techniques (Eurocrypt)},
  year={2023},
  note={ DOI: \href{https://doi.org/10.1007/978-3-031-30589-4_2}{10.1007/978-3-031-30589-4\_2} }
}

@inproceedings{2023_fetzer_universally,
  title={Universally composable auditable surveillance},
  author={Fetzer, Valerie and Kloo{\ss}, Michael and M{\"u}ller-Quade, J{\"o}rn and Raiber, Markus and Rupp, Andy},
  booktitle={International Conference on the Theory and Application of Cryptology and Information Security (Asiacrypt '23)},
  year={2023},
  note={ {DOI: \href{https://doi.org/10.1007/978-981-99-8724-5_14}{10.1007/978-981-99-8724-5\_14}} }
}

@article{2024_hooda_experimental,
  title={Experimental Analyses of the Physical Surveillance Risks in Client-Side Content Scanning},
  author={Hooda, Ashish and Labunets, Andrey and Kohno, Tadayoshi and Fernandes, Earlence},
  year = {2024},
  booktitle = {Proceedings of the Annual Network and Distributed System Security Symposium (NDSS)},
  note={ {DOI: \href{https://doi.org/10.14722/ndss.2024.241401}{10.14722/ndss.2024.241401}} }
}

@inproceedings{2023_jain_deep,
  title={Deep perceptual hashing algorithms with hidden dual purpose: when client-side scanning does facial recognition},
  author={Jain, Shubham and Cre{\c{t}}u, Ana-Maria and Cully, Antoine and de Montjoye, Yves-Alexandre},
  booktitle={IEEE Symposium on Security and Privacy (Oakland '23)},
  year={2023},
  note={ {DOI: \href{https://doi.org/10.1109/SP46215.2023.10179310}{10.1109/SP46215.2023.10179310}} }
}

@article{2024_abelson_bugs,
  title={Bugs in our pockets: The risks of client-side scanning},
  author={Abelson, Harold and Anderson, Ross and Bellovin, Steven M and Benaloh, Josh and Blaze, Matt and Callas, Jon and Diffie, Whitfield and Landau, Susan and Neumann, Peter G and Rivest, Ronald L and others},
  journal={Journal of Cybersecurity},
  volume={10},
  number={1},
  year={2024},
  note={ {DOI: \href{https://doi.org/10.1093/cybsec/tyad020}{10.1093/cybsec/tyad020}} }
}

@inproceedings{2022_jain_adversarial,
  title={Adversarial Detection Avoidance Attacks: Evaluating the robustness of perceptual hashing-based client-side scanning},
  author={Jain, Shubham and Crețu, Ana-Maria and de Montjoye, Yves-Alexandre},
  booktitle={31st USENIX Security Symposium (USENIX Sec '22)},
  year={2022}
}

@inproceedings{1991_pedersen_commitment,
  title={Non-interactive and information-theoretic secure verifiable secret sharing},
  author={Pedersen, Torben Pryds},
  booktitle={Annual international Cryptology Conference (CRYPTO '91)},
  year={1991}
}

@article{2016_alsabah_performance,
  title={{Performance and Security Improvements for Tor: A Survey}},
  author={AlSabah, Mashael and Goldberg, Ian},
  journal={ACM Computing Surveys (CSUR)},
  volume={49},
  number={2},
  year={2016}
}

@article{2022_darir_mleflow,
  title={{MLEFlow: Learning from History to Improve Load Balancing in Tor}},
  author={Darir, Hussein and Sibai, Hussein and Cheng, Chin-Yu and Borisov, Nikita and Dullerud, Geir and Mitra, Sayan},
  journal={Proceedings on Privacy Enhancing Technologies (PETS '22)},
  year={2022}
}

@misc{ip_cert,
	author = {Aaron Gable},
	year = {2025},
	title = {{We've Issued Our First IP Address Certificate}},
	url = {{https://letsencrypt.org/2025/07/01/issuing-our-first-ip-address-certificate/}},
	note = {Last accessed: July 4, 2025},
}

@inproceedings{2006_curtmola_searchable,
  title={{Searchable Symmetric Encryption: Improved Definitions and Efficient Constructions}},
  author={Curtmola, Reza and Garay, Juan and Kamara, Seny and Ostrovsky, Rafail},
  booktitle={Proceedings of the 13th ACM Conference on Computer and Communications Security~(CCS '06)},
  year={2006}
}

\appendix

\section{Integration into Tor's Handshake}
\label{app:extension}

\begin{figure*}[t]
	\centering
	\includegraphics[width=\textwidth]{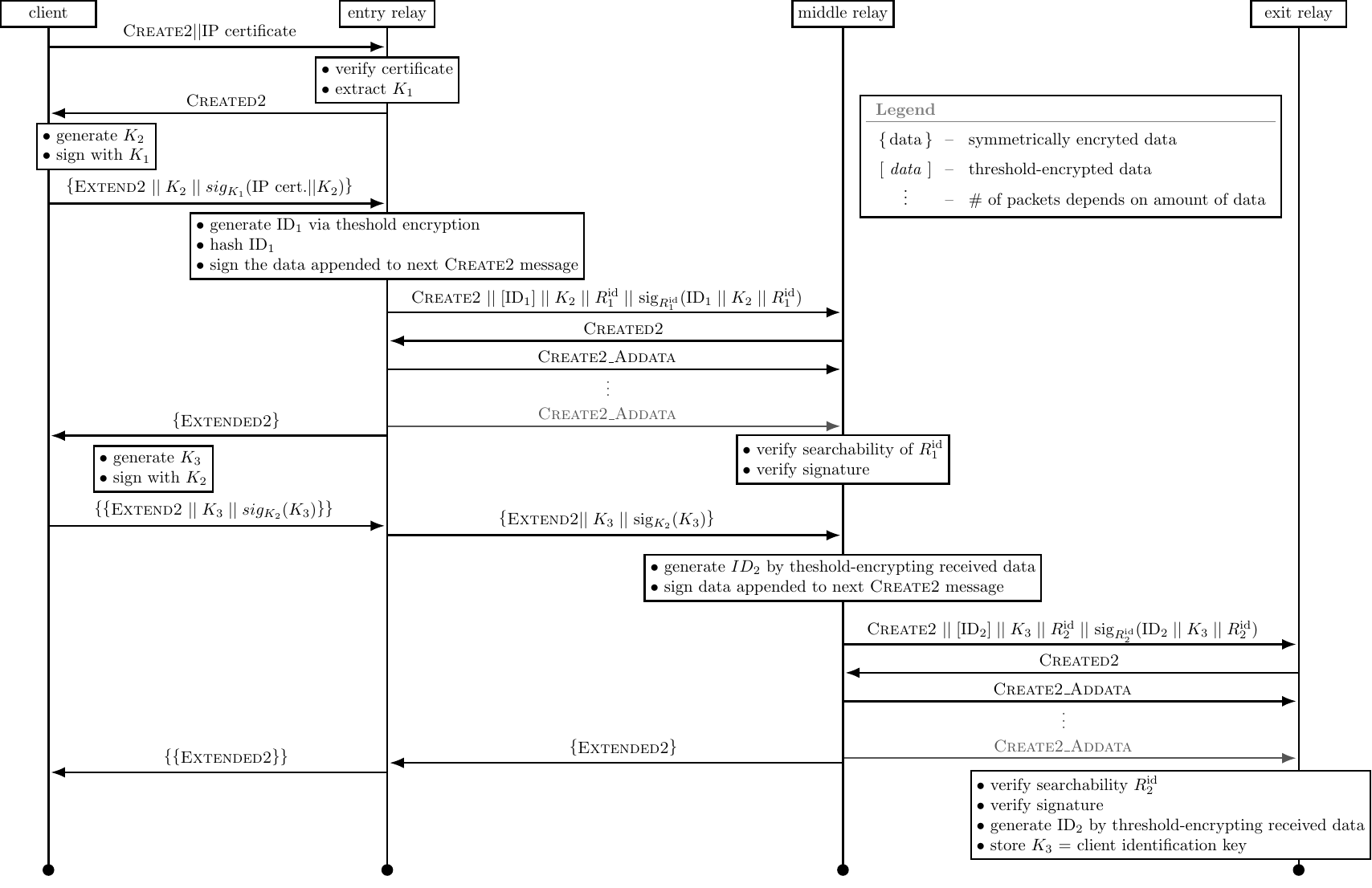}
	\caption{\name's oblivious authentication protocol integrates nicely into Tor's circuit establishment without causing additional round trips.}
	\label{fig:handshake}
\end{figure*}

We now discuss how \name integrates its oblivious authentication into Tor's circuit establishment.
Figure~\ref{fig:handshake} depicts the extensions made to Tor's circuit establishment.
The data exchange of the original Tor handshake is compiled in the cell types, which are indicated in capital letters.
\name extends the relevant cell types, \ie \textsc{Create2}, \textsc{Extend2}, and \textsc{Extended2} cells with the data necessary for the oblivious authentication.
The clients and relays append the IP certificates or encrypted identities to \textsc{Create2} cells.
This data, however, does not always fit in a single cell. 
Therefore, the data is fragmented, and the first fragment is appended to the \textsc{Create2} cell.
All further fragments are sent in the newly added \textsc{Create2\shortunderscore Addata} cells. 
\name dispatches these cells immediately after transmitting \textsc{Create2} cells.
As soon as a relay receives the \textsc{Create2} cell (not necessarily all \textsc{Create2\shortunderscore Addata} cells), it executes the Tor handshake and sends a \textsc{Created2} cell back.
After a relay has sent all \textsc{Create2\shortunderscore Addata} cells, it sends a standard \textsc{Extended2} cell back to the client to inform it about the successful circuit extension.
Once all \textsc{Create2\shortunderscore Addata} have been received by the extending relays, they perform all checks as explained in Section~\ref{sec:authentication}.
Overall, \name's oblivious authentication handshake can thus be integrated seamlessly into Tor's circuit extension without introducing any additional round trips.

\section{Opaquely Retrieving IP Certificates}
\label{app:ipcert}

The simple form of retrieving temporal IP certificates is not feasible if the network operator does not tolerate the usage of \name, \eg how the Chinese Firewall blocks regular Tor traffic~\cite{2018_dunna_torblock}.
Certificate requests are easily interceptable and blockable, or worse, the requester could be prosecuted.
Therefore, we discuss in the following how someone who wishes to use \name could securely request IP certificates under these circumstances.
In the past, much effort was invested into making Tor traffic opaque to traffic analysis.
Similarly to how bridge relays work, CAs and clients can communicate in a way that does not resemble classical certificate requests.
For this, either the CA needs access to enough different IP addresses that not all of them can be blocked simultaneously, or the CA can accept certificate requests through proxies.
The process to request certificates in the latter case works as follows.
Because the CA cannot verify the origin of a request, the host of the proxy server could request certificates for whatever IP address they desire.
Therefore, the certificate itself is not distributed over the same proxy as the request.
Instead, the proxy is only informed about other servers that will host shares of the certificate and communicates this to the requesting client.
The client can then establish covert connections to these servers and retrieve shares of its certificate.
Each of these new proxies verifies the IP address of the client, which ensures that certificates are only created for the user controlling the certified IP address.
A single honest proxy suffices to ensure correctness and also identify misbehavior of other proxies.

\section{Hybrid Threshold Encryption}
\label{app:encryption}

To efficiently implement \name, we designed and implemented a hybrid threshold-encryption scheme based on RSA and AES, which is published as the thRSAhold Python library.\footnote{thRSAhold is available on PyPi and on GitHub at \url{https://github.com/eric-wagner/thRSAhold}.}
Our thRSAhold library is compatible with standard RSA and AES implementations, such that, \eg \emph{openssl} can be used for fast encryption, and the slower Python code is only used for deanonymizations.

Our threshold encryption closely follows Shoup's threshold RSA signatures~\cite{2000_shoup_threshold-rsa}.
All plaintexts are padded to be at least as long as the modulus of the public RSA key to prevent common attacks against RSA.
If the plaintext does not fit into a single RSA ciphertext, a random AES key is generated and prepended to the plaintext.
This key and the beginning of the original plaintext message are threshold-encrypted, while the remaining plaintext is encrypted with AES in GCM mode to ensure confidentiality \emph{and} data authenticity.

During decryption, the first part of the message, clearly defined by the size of the public RSA modulus, is threshold decrypted by the consortium.
Only an entity that collects enough decryption shares can then reconstruct the first part of the plaintext, which only requires knowledge about the public encryption key.
Thus, the requesting \ac{LEA} can decrypt the identity of the suspicious user in the final deanonymization step, without revealing this information to the consortium itself.
If the ciphertext is longer than the RSA modulus, the first \SI{32}{\byte} of the decrypted plaintext are interpreted as an AES key to decrypt the remaining ciphertext. 
The key can then be discarded afterward.
Finally, the padding is removed from the plaintext to reveal the original plaintext.

\section{Consortium Smart Contract}
\label{app:smart-contract}

Listing~\ref{lst:smart-contract} shows an exemplary Solidity smart contract to manage a static consortium of five members with a basic majority vote to accept a deanonymization request. The highlighted lines of code show the variables and events that are revealed to the involved exit node to prove that the deanonymization request for a specific flow record has been accepted and to the public (either in real time or through delayed disclosure).

\begin{lstlisting}[
	language=Solidity,
	breaklines=true,
	postbreak=\mbox{\textcolor{red}{$\hookrightarrow$}\space},
	breakatwhitespace=true,
	caption={A minimal smart contract to manage a fixed consortium of size 5 with an threshold to accept a deanonymization request of 3. The information leaked in real time to the public is highlighted in \lcolorbox{highlight-stats}{\strut orange}, the information revealed to convince exit nodes of a successful vote are highlighted in \lcolorbox{highlight-exit}{\strut purple}, and the additional information revealed during the delayed disclosure to the public is highlighted in \lcolorbox{highlight-public}{\strut green}.},
    escapechar=|,
	label={lst:smart-contract}
]
pragma solidity >=0.8;

contract Voting {

  Case[] private cases;
  uint256 next_case_id=0;
  uint256 constant VOTING_THRESHOLD = 3;
  address[] consortium = [
    0x5B38Da6a701c568545dCfcB03FcB875f56beddC4, 0xAb8483F64d9C6d1EcF9b849Ae677dD3315835cb2,
    0x4B20993Bc481177ec7E8f571ceCaE8A9e22C02db, 0x78731D3Ca6b7E34aC0F824c42a7cC18A495cabaB,
    0x617F2E2fD72FD9D5503197092aC168c91465E7f2
  ];

  event Accepted(uint256 indexed index, bytes32 key);
  event NewCase(uint256 indexed index, bytes32 key);

  struct Case{
    bytes32 hash;
    int256 accepted; // 32 byte such that the variable can be revealed individually
    mapping(address=>bool) votes_by_consortium;
    string reasoning;
  }
    
  function request_deanonymization(bytes32 hash, string memory reasoning) public returns (uint256 id){
    Case storage c = cases.push();
    |\lcolorbox{highlight-exit}{\strut c.hash = hash;}|
    for(uint i=0; i<consortium.length; i++){
        c.votes_by_consortium[consortium[i]] = false;
    }
    |\lcolorbox{highlight-exit}{\strut c.accepted = -1;}|
    |\lcolorbox{highlight-public}{\strut c.reasoning = reasoning;}|
    |\lcolorbox{highlight-stats}{\strut emit NewCase(next\_case\_id, cases[id].hash);}|

    return next_case_id++;
  }
    
  function vote_in_favor(uint256 id) public returns (bool){ 
    |\lcolorbox{highlight-public}{\strut cases[id].votes\_by\_consortium[msg.sender]=true;}| // track all votess
    
    uint256 count = 0;
    for(uint i=0; i<consortium.length; i++){
      if(cases[id].votes_by_consortium[consortium[i]]){ // only count votes by members
        count++;
      }
    }
    
    if(count>=VOTING_THRESHOLD && cases[id].accepted!=0){
      |\lcolorbox{highlight-exit}{\strut cases[id].accepted = 0;}|
      |\lcolorbox{highlight-stats}{\strut emit Accepted( id, cases[id].hash );}|
    }

    return cases[id].accepted==0;
  }
}

\end{lstlisting}

\end{document}